\documentclass[fleqn,usenatbib]{mnras}

\usepackage{newtxtext,newtxmath}

\usepackage[T1]{fontenc}

\DeclareRobustCommand{\VAN}[3]{#2}
\let\VANthebibliography\thebibliography
\def\thebibliography{\DeclareRobustCommand{\VAN}[3]{##3}\VANthebibliography}

\usepackage{graphicx}	
\usepackage{amsmath}	
\usepackage[utf8]{inputenc}
\usepackage{amsmath}
\usepackage{booktabs}
\usepackage{multirow} 
\usepackage{bm}
\usepackage[utf8]{inputenc}
\usepackage[T1]{fontenc}
\usepackage[justification=centering]{caption}
\usepackage[para]{threeparttable}

\title[\texttt{Tempawral}]{\texttt{Tempawral}: A Time-Resolved Retrieval Framework for Variable Brown Dwarfs and Exoplanets}

\author[Fei Wang et al.]{Fei Wang,$^{1}$\thanks{E-mail:f.wang5@herts.ac.uk}
Ben Burningham,$^{1}$
Stuart Littlefair,$^{2}$
Etienne Artigau,$^{3}$
Yuka Fujii,$^{4,5}$
Jacqueline K. Faherty,$^{6,7}$
\newauthor and Johanna M. Vos$^{8,6}$
\\
$^{1}$Department of Physics, Astronomy and Mathematics, University of Hertfordshire, Hatfield, UK\\
$^{2}$Department of Physics and Astronomy, University of Sheffield, Sheffield S3 7RH, UK\\
$^{3}$D\'epartement de Physique, Universit\'e de Montr\'eal, Montr\'eal, Canada\\
$^{4}$Division of Science, National Astronomical Observatory of Japan, 2-21-1 Osawa, Mitaka, Tokyo 181-8588, Japan\\
$^{5}$Graduate Institute for Advanced Studies, SOKENDAI, 2-21-1 Osawa, Mitaka, Tokyo 181-8588, Japan\\
$^{6}$Department of Astrophysics, American Museum of Natural History, Central Park West at 79th Street, New York, NY 10034, USA\\
$^{7}$Department of Physics, The Graduate Center City University
of New York, New York, 10016, NY, USA\\
$^{8}$ School of Physics, Trinity College Dublin, The University of Dublin, Dublin, Ireland
}

\date{Accepted XXX. Received YYY; in original form ZZZ}

\pubyear{\the\year{}}

\begin{document}
\label{firstpage}
\pagerange{\pageref{firstpage}--\pageref{lastpage}}
\maketitle

\begin{abstract}
Brown dwarfs and exoplanets are thought to host complex atmospheric phenomena such as clouds, storms, and chemical heterogeneity, akin to weather patterns on Earth. These features can produce pronounced spectral variability. Time-variability monitoring provides a unique window into surface inhomogeneities that cannot be directly resolved with foreseeable imaging technology. Current time-series analysis techniques have provided qualitative constraints on variability mechanisms but lack the ability to quantitatively estimate the extent of variation on atmospheric properties. We present \texttt{Tempawral}, the first data-driven time-resolved atmospheric retrieval framework that quantitatively retrieving variability in atmospheric parameters via an eigen-spectra inversion technique, leveraging the full spectra dataset. We validate this method on simulated time-series spectra of a variable brown dwarf, demonstrating that it successfully recovers key variability drivers, including inhomogeneous cloud coverage, evolving chemical abundances, and changes in temperature structure. We further showcase the utility of \texttt{Tempawral} by applying it to JWST/NIRISS-SOSS time-series observations of a highly variable T2.5 brown dwarf. The observed variability is best explained by a $\sim$300 K temperature perturbation near the 1-bar, accompanied by variations in the abundances of $\rm H_{2}O$, $\rm CO$, $\rm FeH$, as well as changes in the thickness of the iron cloud deck. This work provides a generalized framework for time-resolved atmospheric retrievals in the JWST era, enabling comprehensive interpretations of dynamic atmospheric processes in substellar objects.
\end{abstract}



\section{Introduction}\label{sec:intro}


Brown dwarfs (BDs) represent a crucial link between low-mass stars and giant planets. Their atmospheric diversity is primarily governed by effective temperature ($T_{\rm eff}$), surface gravity, and metallicity, parameters that span a wide range overlapping with both cool stars and giant exoplanets. At the hottest end of the brown dwarf sequence ($ T_{\rm eff} \le 2700 K$), their atmospheres resemble those of cool stars and ultra-hot Jupiters, dominated by atomic and ionic species, with significant contributions from continuum and atomic line opacities \citep{allard1997model,lodders2002atmospheric}. As $T_{\rm eff}$ decreases, molecular opacities such as from $\rm H_{2}O$, $\rm CH_{4}$, CO, and $\rm NH_{3}$ become increasingly important. At these cooler temperatures, condensate clouds and disequilibrium chemistry driven by vertical mixing also become prevalent, further shaping the observed spectra and making them more reminiscent of giant exoplanets \citep{FegleyLodders1996, Visscher2006,Coulter2022, Beiler2024}. At the coolest extremes, Y dwarfs with temperatures as low as $\sim$ 200K nearly overlap with the temperature regime of solar system gas giants. These key processes of cloud formation, molecular absorption, and chemical disequilibrium remain poorly understood in planetary atmospheres. BDs offer data-rich laboratories for studying the many of atmospheric processes that are central to our understanding of giant planets.

Like the gas giants in our Solar System, most brown dwarfs are rapid rotators, with rotation periods typically ranging from 1 to 20 hours \citep{osorio2006spectroscopic,snellen2014fast,bryan2020worlds,tannock2021weather,vos2022let}. Time-resolved light curve observations in the optical, near-infrared, and mid-infrared have revealed dramatic and ubiquitous variability across nearly all spectral types \citep{apai2013hst,radigan2014independent,metchev2015weather,apai2017zones,manjavacas2017cloud,lew2020cloud,zhou2020spectral,vos2022let,cushing2016first,leggett2016observed}. These studies suggest that the majority of L and T dwarfs are variable, with typical amplitudes exceeding 0.2\%. Large variability surveys \citep{radigan2014independent,metchev2015weather}, which examined samples of 62 and 44 brown dwarfs respectively, further show that objects in the L/T transition exhibit stronger and higher amplitude variability than those outside the transition \citep{radigan2014independent,eriksson2019detection}, with amplitudes ranging from a few percent up to 26\%. Several physical mechanisms has been proposed to explain this variability, including inhomogeneous cloud cover as a dominant driver in the L/T transition \citep{radigan2014independent}, as well as localized temperature perturbations or hotspots \citep{robinson2014temperature}, thermochemical instabilities \citep{tremblin2016cloudless}, and auroral activity \citep{hallinan2015magnetospherically}. These observed variability patterns are thought to arise from underlying atmospheric dynamics, such as cloud radiative feedback \citep{tan2021atmospheric} and convective perturbations \citep{zhang2014atmospheric}.

Constraining the origins of the variability was limited by incomplete observational datasets. Traditional photometric monitoring suffers from limitations in atmospheric depth sensitivity due to the use of only a few broadband filters. Multi-wavelength, phase-resolved observations can provide a comprehensive three-dimensional view of the atmosphere and help disentangle the mechanisms that drive variability across different pressure levels \citep{morley2014spectral,biller2024jwst,mccarthy2025jwst}. 
JWST enables time-resolved spectroscopy from the visible to the mid-infrared, with NIRSpec covering 0.6–5 $\mu$m and MIRI extending to 28 $\mu$m \citep{jakobsen2022near}. This capability has opened a new window on spectroscopic variability studies of brown dwarfs and planetary-mass objects. \citet{biller2024jwst} presented the first JWST spectroscopic monitoring of the brown dwarf binary WISE 1049AB \citep{luhman2013discovery}, detecting variability from 1–14 $\mu$m, while \citet{mccarthy2025jwst} reported monitoring of the isolated planetary-mass object SIMP 0136 \citep{artigau2006discovery,gagne2017simp} over 0.8–11 $\mu$m. More recently, \citet{akhmetshyn2025mapping} presented JWST NIRISS-SOSS time-series spectroscopy of SIMP 0136, which is also the dataset used in this paper. To date, JWST observing cycles have allocated over 150 hours to time-series studies of 11 brown dwarfs and planetary-mass objects. These targets span the temperature range between 250 and 2300 K and are being observed with time-series spectroscopy to explore the dynamic processes in brown dwarf atmospheres.


Previous techniques for analyzing time-series spectroscopy have primarily focused on methods such as Doppler imaging to produce spherically deconvolved surface maps, light-curve clustering to associate different variability patterns with varying atmospheric pressure levels, and comparing retrievals between selected epochs within the spectral series. The first successful Doppler imaging attempt on a brown dwarf was conducted by \citet{crossfield2014global} to create a surface map of WISE\,1049B. Subsequently, \citet{chen2024global} applied Doppler imaging to WISE\,1049AB using time-resolved, high-resolution spectroscopic observations from Gemini IGRINS. Their analysis revealed persistent spot-like structures located from the equatorial to mid-latitude regions, as well as newly identified polar spots in the $H$- and $K$-band maps. \citet{akhmetshyn2025mapping} tested the sinusoidal and spherical harmonic mapping techniques to extract brightness maps from the light curves at different wavelengths of SIMP\,0136, finding that the spectro-photometric variability connects three distinct spectral regions to three atmospheric layers. The overall variability is highly correlated with changes in temperature, cloud coverage, and possibly effective metallicity.
Doppler imaging successfully recovered the evolving surface features responsible for the spectral variability, while the deconvolution can suffer from non-uniqueness.

The light-curve clustering technique converts time-series spectra into wavelength-dependent light curves and applies k-means clustering (via \texttt{scikit-learn}) to identify distinct variability behaviors. These clusters are mapped onto contribution functions (tells where in the atmosphere the emergent radiation at a given wavelength is coming from) to associate variability with atmospheric pressure levels. Applied to JWST/NIRSpec and MIRI observations of WISE 1049AB, this method revealed variability driven by patchy clouds in deep layers and temperature- or chemistry-induced hotspots at higher altitudes \citep{biller2024jwst,chen2025jwst}. Similarly, \citet{mccarthy2025jwst} applied this technique to JWST spectroscopy of SIMP~0136 (0.8–11 $\mu$m), identifying contributions from clouds, hotspots, and variable carbon chemistry. This approach effectively links the observed variability to atmospheric processes, but it remains qualitative and cannot fully constrain the amplitudes of individual atmospheric variations

A more direct approach is to perform atmospheric retrievals on individual time-step spectra and compare the inferred atmospheric properties. However, this approach is challenged by the high computational cost and the parameter degeneracies in retrieval analyses, particularly when dealing with hundreds to thousands of spectra for a single object (e.g., SIMP~0136, for which $\sim$6000 JWST/NIRSpec PRISM spectra were obtained in GO 3548, PI: Vos). \citet{nasedkin2025jwst} reduced this burden by performing retrievals on 24 phase-binned spectra of SIMP~0136, sampled every 15$^{\circ}$ in rotational phase. Their results suggest that the observed variability may be driven by changes in effective temperature and by variations in the abundances of $\rm CO_{2}$ and $\rm H_{2}S$. However, interpreting variability in terms of effective temperature implicitly assumes changes in the object’s luminosity, and the physical origin of such variations remains uncertain. More generally, the interplay between chemistry, cloud properties, and temperature structure introduces degeneracies that complicate the unique identification of variability drivers.

\begin{figure}
    \centering
	\includegraphics[width=0.9\columnwidth]{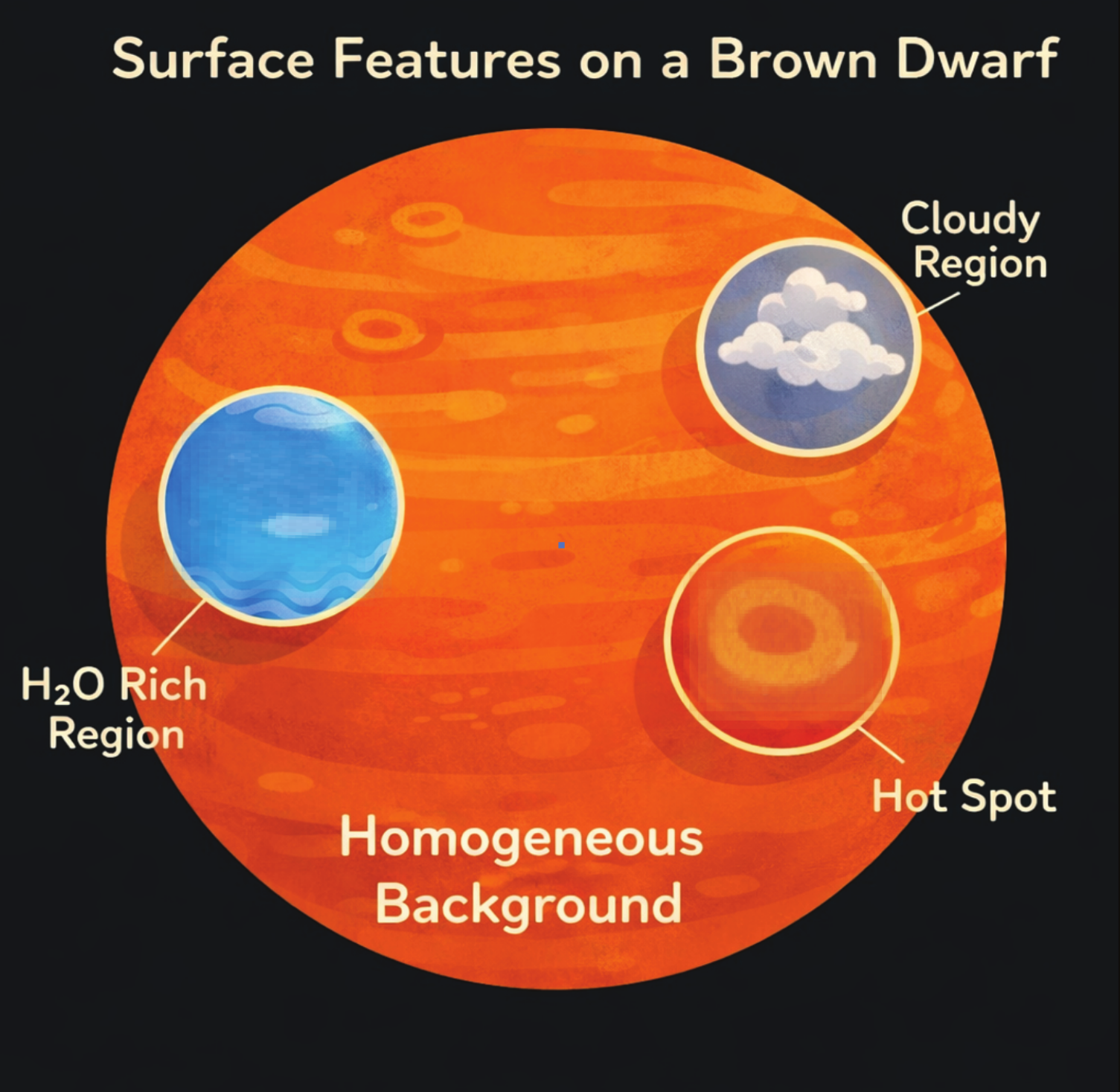}
    \caption{Simplified physical picture for modeling time-variable brown dwarf atmospheres.  Multiple surface regions with distinct chemical compositions, cloud properties, or thermal structures may coexist, evolve, and interact dynamically. When these heterogeneous atmospheric structures are superimposed on a more homogeneous background and rotate in and out of view, the resulting disk-averaged atmospheric properties can exhibit approximately sinusoidal temporal  behavior, which is also the assumption used in \texttt{Tempawral}. However, such atmospheric variability evolution pattern can be more complex (see detailed discussion in Section ~\ref{sec:Sinusoidal}).
\label{fig:surface_features}}
\end{figure}

The above efforts have provided valuable insights into the origins of spectral variability, but they often lack quantitative constraints on the amplitudes of variations in atmospheric properties. With the increasing volume of high quality JWST time series observations, where a single object may yield hundreds to thousands of high signal to noise spectra, there is a growing need for a generalized and computationally efficient time resolved retrieval framework that can fully exploit these rich datasets and quantitatively investigate the physical drivers of spectral variability. To address this need, we developed \texttt{Tempawral}, a data-driven, time-resolved retrieval framework. \texttt{Tempawral} models evolving brown dwarf atmospheres by allowing multiple surface regions with distinct chemical compositions, cloud properties, and thermal structures to emerge, evolve, and interact dynamically (see Figure~\ref{fig:surface_features}), producing complex and spatially correlated surface features \citep{lee2023dynamically, lee2024dynamically} . These heterogeneous atmospheric structures are superimposed on a more homogeneous background and rotate in and out of view, causing disk-averaged atmospheric properties such as gas abundances or longitudinally averaged temperature profiles can follow parameterized temporal behaviors. Different combinations of temporal evolution across atmospheric parameters generate distinct time-series spectra and, consequently, unique principal components (hereafter eigen-spectra) that capture the underlying variability patterns.

\texttt{Tempawral} is designed to recover the temporal evolution of individual atmospheric parameters by performing retrievals directly on eigen-spectra rather than on individual time-series spectra. By operating in eigen-space, \texttt{Tempawral} makes most use of the variability information contained within time series spectra data while substantially improving computational efficiency. Related approaches have applied principal component analysis (PCA) to extract dominant eigencurves from spectroscopic secondary-eclipse light curves for three-dimensional eclipse mapping \citep{mansfield2020eigenspectra}. In contrast, our framework operates in the spectral domain and uses PCA to identify the dominant modes of spectral variability while explicitly incorporating time as a fundamental dimension of atmospheric retrieval. This enables generalized atmospheric fits to variability patterns rather than to the original time-series spectra.


Section ~\ref{sec:data} describes the SIMP\,0136 data used in this paper. In Section~\ref{sec:tempawral}, we provide an overview of the \texttt{Tempawral} code and outline the methodology employed in this study. In Section~\ref{sec:demo}, we outline simulated time-series observations and eigen-spectra retrieval model setup for a SIMP\,0136-like brown dwarf.
Section~\ref{sec:result} presents the results of model demonstrations. In Section~\ref{sec:application}, we apply this method to JWST/NIRISS-SOSS near-infrared time-series data of SIMP\,0136. Section~\ref{sec:dis} remarks on the parametrization of time-series models and highlights caveats related to eigen-spectra retrieval techniques, including the selection of baseline retrievals and the time distribution of variables. Finally, Section~\ref{sec:sum} summarizes the main conclusions of this work and briefly discusses possible directions for future studies.

\renewcommand{\arraystretch}{1.35}
\begin{table}
\centering
\begin{threeparttable}
\caption{Baseline Model Parameters and Retrieved Median Values} 
\begin{tabular}{llrr}
\hline
\hline
\textbf{Category} & \textbf{Parameter} & \textbf{MLE Value\tnote{a}} & \textbf{Median Value} \\
\hline
\multirow{7}{*}{Gas Abundance\tnote{b} } 
  & $\rm H_{2}O$        & -3.82  & $-3.81^{+0.05}_{-0.03}$ \\
  & $\rm CO$            & -3.39 & $-3.37^{+0.05}_{-0.05}$ \\
  & $\rm CH_{4}$        & -4.32 & $-4.32^{+0.05}_{-0.04}$ \\
  & $\rm NH_{3}$        & -5.27 & $-5.31^{+0.05}_{-0.08}$ \\
  & $\rm CrH$          & -9.35 & $-9.39^{+0.09}_{-0.09}$ \\
  & $\rm FeH$          & -9.36 & $-9.32^{+0.05}_{-0.08}$ \\
  & $\rm K\_Na$\tnote{c}          & -6.58 & $-6.52^{+0.07}_{-0.06}$ \\
\hline
\multirow{4}{*}{Refinement parameter\tnote{d}} 
  & $\bm{\log g}$\tnote{e}              & 4.94 & $4.91^{+0.05}_{-0.06}$ \\
  & $\boldsymbol{\mathit{R^{2}/D^{2}}}$\tnote{e}      & $2.37 e^{-19}$  &  $0.00^{+0.00}_{-0.00}$ \\
  & $\Delta\lambda$         & -0.0012 & $-0.00^{+0.00}_{-0.00}$ \\
  & $\bm{\log f}$\tnote{e}    & -30.99 & $-30.97^{+0.02}_{-0.02}$ \\
\hline
\multirow{5}{*}{P-T profile\tnote{f}} 
  & $\alpha_{1}$               & 0.94 & $0.94^{+0.04}_{-0.07}$ \\
  & $\alpha_{2}$               & 0.06 & $0.07^{+0.03}_{-0.01}$ \\
  & $ \log P_{1}$               & 0.21 & $0.09^{+0.09}_{-0.32}$ \\
  & $\log P_{3}$               & 1.58 & $1.66^{+0.20}_{-0.11}$ \\
  & $ T_{3}$                    & 3889.83 & $3799.47^{+132.43}_{-169.25}$ \\
\hline
\multirow{10}{*}{Cloud\tnote{g}} 
  & $ f_{\rm cld}$              & 0.91 & $0.89^{+0.01}_{-0.06}$ \\
  & \multicolumn{2}{l}{\underline{ $\rm Mg_{2}SiO_{4}$ slab cloud}} \\
  & \quad $\rm \tau_{\rm mcs}$     & 25.85 & $24.61^{+1.21}_{-1.76}$ \\
  & \quad $\log p_{\rm mcs}$   & -3.48 & $-1.08^{+0.88}_{-2.07}$ \\
  & \quad $dp_{\rm mcs}$ & 4.14 & $1.70^{+2.09}_{-0.93}$ \\
  & \quad $\bm{ \mathbf{a_{\rm mcs}}}$\tnote{e}      & -0.84 & $-0.87^{+0.13}_{-0.24}$ \\
  & \quad $\bm{\mathbf{b_{\rm mcs}}}$\tnote{e}        & 0.21 & $0.19^{+0.15}_{-0.12}$ \\
  & \multicolumn{2}{l}{\underline{Fe deck cloud}} \\
  & \quad $\log p_{\rm mcd}$   & 1.03 & $1.03^{+0.07}_{-0.04}$ \\
  & \quad $dp_{\rm mcd}$ & 1.08 & $3.57^{+2.25}_{-1.80}$  \\
  & \quad $\bm{\mathbf{a_{\rm mcd}}}$\tnote{e}      & -1.01 & $-1.02^{+0.11}_{-0.89}$ \\
  & \quad $\bm{\mathbf{b_{\rm mcd}}}$\tnote{e}      & 0.66 & $0.55^{+0.29}_{-0.29}$ \\
\hline
\end{tabular}
\begin{tablenotes}
\item[a] MLE Value denotes maximum likelihood estimate value.

\item[b] Gas abundance is presumably
$\log_{10}$ of a vertically-uniform mixing ratio.

\item[c] $\rm K\_Na$ is combined Na and K volume mixing ratio, the ratios of K and Na are tied to the solar ratio. 

\item[d] $\bm{\log g}$ is in cgs units; $\Delta\lambda$ is in microns; $\bm{\log f}$ is tolerance factor. 

\item[e] Bold parameters like $\bm{\log g}$, $\boldsymbol{\mathit{R^{2}/D^{2}}}$, $\bm{\log f}$ (tolerance factor) and cloud particle size distributions are assumed to be invariant in the subsequent time-resolved retrievals. 

\item[f] The atmospheric temperature profile is parameterized following \citet{madhusudhan2009temperature}.

\item[g] The cloud parametrization follows \citet{burningham2021cloud}.
\end{tablenotes}

\label{tab1}
\end{threeparttable}
\end{table}

\section{SIMP\,0136 Time-resolved Spectra} \label{sec:data}

The data for this research were obtained with the Near Infrared Imager and Slitless Spectrograph in the Single Object Slitless Spectroscopy mode (NIRISS/SOSS) onboard JWST as a part of cycle 1 Guaranteed Time Observations  program \#1209 (PI: Artigau), The observation detail can be found in \citet{akhmetshyn2025mapping}. The SOSS-Inspired SpectroScopic Extraction pipeline is used for data reduction \citep{Lim2023}. The reduced data consists of 81 spectra ($R\sim$1200) from 0.85 to 2.83\,$\mu$m taken over slightly more than one rotational period.
SIMP~0136 is an isolated planetary-mass object that has been extensively studied using both ground-based and space-based observations \citep{artigau2006discovery,artigau2009photometric,apai2013hst,yang2016extrasolar,mccarthy2024multiple}. With a spectral type of \hbox{T$2.5\pm0.5$} \citep{artigau2006discovery} and an effective temperature $1360 \pm 20$\,K \citep{vos2023patchy}, SIMP~0136 lies in the L/T transition region, known for high-amplitude variability likely caused by the variations in the cloud patchiness \citep{radigan2014independent,liu2024near}. Observed variability amplitudes reach up to 5\% in the $J$ band \citep{artigau2009photometric}, and the object has a rotation period of approximately 2.4\,hours \citep{croll2016long,yang2016extrasolar}. Near-infrared variability has been linked to patchy clouds \citep{apai2013hst,mccarthy2024multiple}, while mid-infrared variability may result from $\rm CO$/$\rm CH_{4}$  fingering convection \citep{tremblin2020rotational}. In addition, strong pulsed radio emission from SIMP\,0136 suggests auroral activity \citep{kao2016auroral,kao2018strongest}.

Previous retrieval analyses (e.g., \citealp{vos2023patchy}) have shown that the 1–15 $\rm \mu m$ spectrum of SIMP\,0136 is best explained by a two-layer cloud structure: a high-altitude forsterite (Mg$_{2}$SiO$_{4}$) cloud overlaying a deeper, optically thick iron (Fe) cloud. In our \texttt{Tempawral} analysis, we follow this same cloud prescription. The atmospheric temperature profile is parameterized following \citet{madhusudhan2009temperature}, and we perform retrievals on the time-averaged spectrum. The resulting best-fit parameters are summarized in Table~\ref{tab1}. The spectrum is best explained by absorption from H$_{2}$O, CO, CO$_{2}$, CH$_{4}$, NH$_{3}$, CrH, FeH, Na, K, along with a pressure–temperature profile that is isothermal at high altitudes and adopts an adiabatic-sloped structure below 10\,bars. The cloud structure follows the two-layer prescription described above with a patchy, high-altitude forsterite cloud overlaying a deeper, optically thick iron cloud.

\begin{figure*}
    \centering
	\includegraphics[width=1.9\columnwidth]{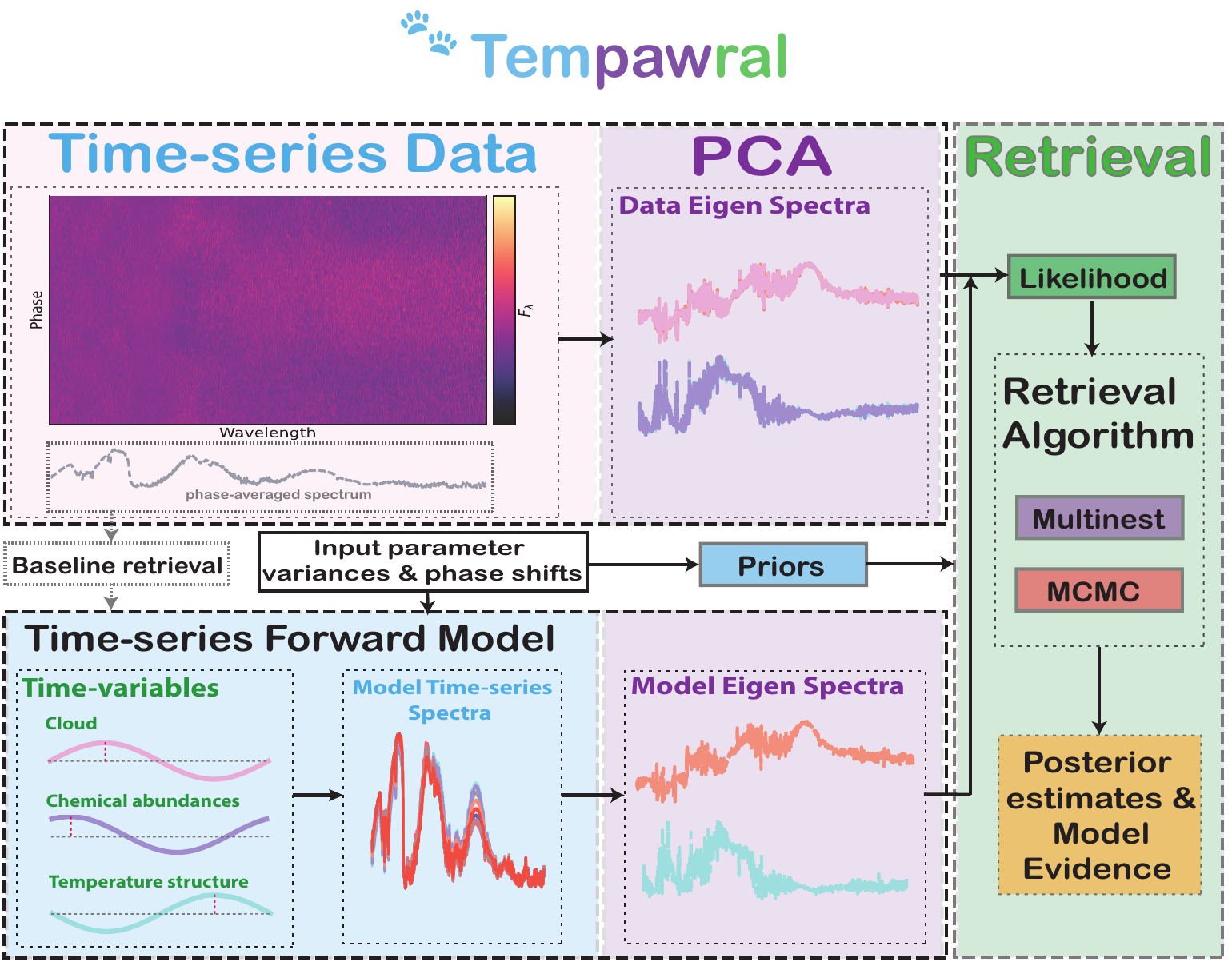}
    \caption{Schematic overview of the \texttt{Tempawral} framework. Starting from a time-series spectroscopic dataset, a baseline atmospheric retrieval is performed on the time-averaged spectrum to obtain reference values on model parameters. Time-variable parameters (temperature, chemical abundances, and cloud properties) are perturbed sinusoidally to generate the time-evolution model parameters; which is fed into the time-series forward model to produce simulated spectra. PCA is then applied to both observed and simulated time-series spectra to extract eigen-spectra, capturing the dominant modes of wavelength-dependent variability. The Bayesian retrieval module compares these eigen-spectra, inferring posterior distributions for the perturbation amplitudes and phase shifts of the atmospheric parameters.
\label{fig:Tflowchart}}
\end{figure*}

\section{\texttt{Tempawral} framework} \label{sec:tempawral}

\texttt{Tempawral}\footnote{\url{https://github.com/fwang23nh/Tempawral}} is designed to quantitatively probe the origins of variability in time-series spectra. Traditional spectral retrieval approaches typically perform independent retrievals on each spectrum in a time series, which becomes computationally prohibitive for objects with thousands of spectra. Instead, \texttt{Tempawral} focuses directly on the variability patterns present in the time-series data, representing these patterns through principal component analysis and comparing them between observed and simulated datasets. Unlike conventional retrieval frameworks, which are generally built around a single-epoch (hereafter `snapshot') forward model coupled with a retrieval algorithm, \texttt{Tempawral} consists of three main components: Time-series spectral generation module, which simulates spectra across selected time steps; PCA-based post-processing module, which reduces data dimensionality and captures the dominant modes of variability; and Bayesian retrieval module, which performs parameter inference directly on dominant modes of variability. Figure~\ref{fig:Tflowchart} presents a schematic overview of the \texttt{Tempawral} workflow. In the following subsections, we describe each of these components in detail, with particular emphasis on the time-series forward model and the PCA-based post-processing step. We then present the complete \texttt{Tempawral} workflow, illustrating how these elements work together to quantitatively characterize atmospheric variability.

\subsection{Time-series Forward Model}

The time-series forward model generates synthetic spectra that evolve over time in response to physically motivated perturbations in the atmospheric state. Starting from a baseline atmospheric model, typically obtained from a retrieval on the time-averaged spectrum, the model introduces time dependent variations in selected parameters such as temperature, chemical abundances, or cloud properties. Figure~\ref{fig:Tmodelflowchart} provides a schematic overview of the time-series forward model.

\subsubsection{Snapshot Forward Model}

The time-series forward model in \texttt{Tempawral} builds upon the snapshot forward model implemented in the \texttt{Brewster\_v2}, which will be released soon, and is an upgraded version of the \texttt{Brewster} retrieval framework \citep{burningham2017retrieval,burningham2021cloud}. The time-series model extends the snapshot model to allow selected atmospheric parameters to vary with time. Readers are referred to \citet{burningham2017retrieval} for a full description of the snapshot framework. Here, we summarize the essential components.

The snapshot forward model computes the emergent emission spectrum under the assumption of plane parallel geometry, using a parametric pressure temperature ($P$-$T$) profile and prescribed chemical abundances. It includes opacity from atomic and molecular lines, Rayleigh scattering, and collision induced absorption (CIA), and incorporates a treatment of patchy clouds using multiple atmospheric columns \citep{vos2023patchy}. Radiative transfer is solved independently for each column, and the resulting spectra are linearly combined according to their fractional coverage to produce the total emergent flux. Using this foundation, the time-series forward model can generate spectra at any desired time step once the time dependent values of all varying parameters are specified.

\subsubsection{Reference Values from Baseline Retrieval}

Given a time-series spectroscopic dataset for a brown dwarf, we first determine the baseline atmospheric state by performing a retrieval on the time-averaged spectrum. The time-averaged spectrum is constructed by stacking hundreds of time-resolved spectra, effectively smoothing out short-term variability. The time-averaged spectrum is assumed to represent the underlying baseline atmosphere, while the observed spectral variability is attributed to time-dependent surface features modulated by rotation. By exploring combinations of temperature structures, cloud models, and chemical compositions, we identify the best-fitting atmospheric model that reproduces the mean spectral shape. The resulting maximum likelihood estimates on each model parameters define the reference parameter set for all model parameters.

\subsubsection{Time-Variable Parameters}\label{subsec:timeV_params}

Spectral variability arises from changing atmospheric properties as the object rotates. The chemistry, cloud structure, and temperature profile within a surface feature may all differ from the surrounding atmosphere, creating a complex heterogeneous pattern that evolves with rotation. As these features rotate into and out of view, their contribution to the disk-integrated spectrum produces periodic spectral variability.

In \texttt{Tempawral}, this behavior is parameterized using three main classes of time-variable parameters:

\begin{enumerate}
\item[\textbf{1. Chemical Abundances.}]
All retrieved gas species are initially treated as time-variables. Each species is assigned a perturbation amplitude and a phase shift. This approach captures the possibility that localized chemical enrichment or depletion, driven by vertical mixing, convection, or photochemical processes, may contribute to the observed spectral changes.

\item[\textbf{2. Cloud Properties.}]
The baseline cloud structure is taken from the best-fit of time-averaged spectrum retrieval and may include slab or deck clouds \citep{burningham2017retrieval} at different pressure levels, composed of either Mie particles or power-law/gray species. The time-variable cloud parameters thus include vertical and horizontal distributions (cloud-top, cloud-base pressures and patchy cloud coverage) and slab cloud optical thickness. Although all cloud parameters could, in principle, vary with rotation, we fix the cloud scattering properties over time to reduce model complexity and mitigate degeneracies. The power-law index of the optical thickness and the Mie particle sizes remain fixed at their baseline values.

\item[\textbf{3. Temperature Structure.}]
The baseline pressure temperature  profile is obtained from the time averaged retrieval. Instead of perturbing individual $P$-$T$ parameters, \texttt{Tempawral} applies a temperature perturbation directly to the baseline profile. This perturbation represents localized thermal features, such as hot or cold spots caused by non-equilibrium chemistry, auroral activity, or radiative cooling. Three parameters define the perturbation: a reference pressure $\log p_{\rm ref}$ indicating the center of the perturbation, a pressure width $\log dp$ describing its vertical extent, and an amplitude $T_{\rm var}$ specifying the temperature deviation.
\end{enumerate}

\subsubsection{Time Evolution of Model Parameters}

Once the time-variable parameters are defined, we simulate their temporal evolution. Given an observed time-series spectral dataset, the initial estimates of the atmospheric properties are obtain from the baseline retrievals on the time averaged spectrum. 
Here we use maximum likelihood estimates on each model parameters as reference values. Starting from these reference values that represent the atmospheric properties of the homogeneous background ( see Figure~\ref{fig:surface_features}), we assume that the observed variability arises from rotational modulation of multiple surface features (time-variable parameters).

Each parameter is assumed to vary sinusoidally about its reference value, reflecting periodic modulation caused by rotating surface features. To allow for asynchronous variability among different atmospheric processes, each time-variable parameter is assigned an independent phase shift. These perturbations are described by

\begin{equation}
m(t) = m_{\text{ref}} + m_{\text{var}} \sin(t + m_{\text{shift}}),
\label{euq1}
\end{equation}

where \(t\) is the rotational phase, expressed as $t = \left(\frac{T}{P_{\rm rot}}\right) 2\pi$, with \(T\) denoting a specific time and $P_{\rm rot}$ the rotational period of the object determined from prior observations. Here, \(m(t)\) is the phase-dependent value of the parameter, \(m_{\rm ref}\) is the reference value from the baseline retrieval, \(m_{\rm var}\) is the perturbation amplitude, and \(m_{\rm shift}\) is the phase shift. Each time-variable parameter therefore follows its own sinusoidal trajectory, while fixed parameters remain constant at their reference values.

\begin{figure}
	\includegraphics[width=0.99\columnwidth]{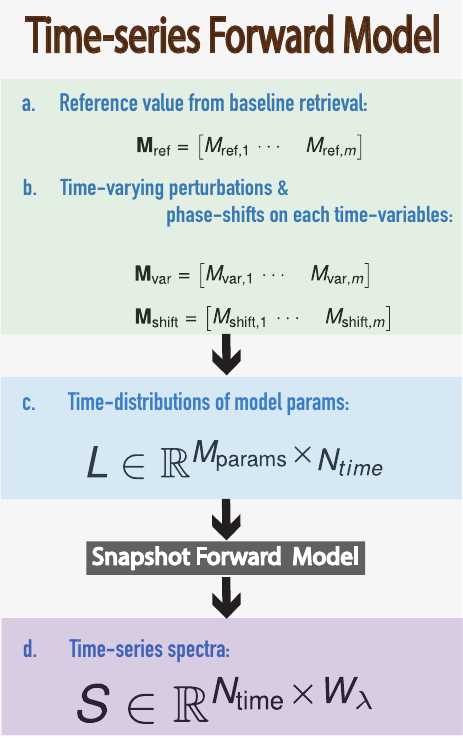}
    \caption{Flowchart of the \texttt{Tempawral}'s time-series forward model. The baseline model provides a reference matrix, $\mathbf{M}_{\rm ref}$, derived from the best-fit solution to the time-averaged spectrum of the observed object. Periodic modulations from rotating atmospheric features are captured through variability amplitude and phase-shift terms, yielding the perturbation and phase-shift matrices, $\mathbf{M}_{\rm var}$ and $\mathbf{M}_{\rm shift}$, which are sampled at each retrieval iteration. The resulting matrix, $\mathbf{L}$, encodes the full time evolution of all model parameters, with each column representing a complete parameter set at a given time step. These parameter sets are passed to the snapshot forward model to generate the simulated time-series spectra, $\mathbf{S}$.}
    \label{fig:Tmodelflowchart}
\end{figure}

As illustrated in Figure~\ref{fig:Tmodelflowchart}, for a dataset with $N$ time steps (modeled using $N$ phase steps), this procedure produces an $\mathbf{L}$ matrix describing the time evolution of all $M$ model parameters:

\begin{equation}
\mathbf{L} =
\begin{bmatrix}
\boxed{L_{11}} & L_{12} & \cdots & L_{1N} \\
\boxed{L_{21}} & L_{22} & \cdots & L_{2N} \\
\vdots         & \vdots & \ddots & \vdots \\
\boxed{L_{M1}} & L_{M2} & \cdots & L_{MN}
\end{bmatrix}_{\substack{\text{M Parameters} \\ \downarrow}}^{\overset{\text{N Time Steps}}{\rightarrow}}
\label{eq:Lmatrix}
\end{equation}

with
\[
L_{ij} = M_{\mathrm{ref},i} + M_{\mathrm{var},i} \sin(t_j + M_{\mathrm{shift},i}).
\]

where $M_{\mathrm{ref}}$ is the matrix of reference values obtained from the retrieval on the time-averaged spectrum, while $M_{\mathrm{var}}$ and $M_{\mathrm{shift}}$ represent the matrices of time-variable perturbation amplitudes and phase shifts for the model parameters, respectively. The model parameters time-evolving matrix L encodes the temporal behavior of all time-variable parameters and serves as input for generating the corresponding time-series spectra in the forward model.

\subsubsection{Simulated Time-series Spectra}

Each column of $\mathbf{L}$ represents the full set of atmospheric parameters at a given time step. These parameter sets are passed to the snapshot forward model to generate the synthetic time-series spectra matrix $\mathbf{S}$, which contains $N$ time steps and $W$ wavelength points:

\begin{equation}
\mathbf{S} =
\begin{bmatrix}
\boxed{S_{11}} & \boxed{S_{12}} & \cdots & \boxed{S_{1W}} \\
S_{21} & S_{22} & \cdots & S_{2W} \\
\vdots & \vdots & \ddots & \vdots \\
S_{N1} & S_{N2} & \cdots & S_{NW}
\end{bmatrix}_{\substack{\text{N Time Steps} \\ \downarrow}}^{\overset{\text{W Wavelengths}}{\rightarrow}}
\label{eq:Smatrix}
\end{equation}

The simulated time-series spectra $\mathbf{S}$ capture both the baseline atmospheric state and the imposed perturbations. These spectra serve as the model predictions that are compared with the observed time-series data in the PCA analysis stage.

\subsection{PCA Post-processing}
\label{subsec:PCA}

Principal Component Analysis is a dimensionality reduction technique that projects high-dimensional data onto a lower-dimensional space while preserving the dominant sources of variance. Applied to time-series spectra, PCA identifies the principal components, which represent orthogonal modes of wavelength-dependent variability that captures distinct spectral variability patterns.

Let \( S \in \mathbb{R}^{N_{\mathrm{time}} \times W_{\lambda}} \) denote the matrix of time-series spectra, where each row corresponds to a time step and each column corresponds to a wavelength channel. Before applying PCA, we subtract the temporal mean spectrum to isolate variability about the mean:

\begin{equation}
\tilde{S}(t,\lambda) = S(t,\lambda) - \langle S(\cdot,\lambda)\rangle_{\mathrm{time}}.
\end{equation}

We then construct the spectral covariance matrix
\begin{equation}
\mathbf{C} = \frac{1}{N_{\mathrm{time}} - 1}\,\tilde{S}^{\top}\tilde{S},
\end{equation}
and solve the eigenvalue problem:
\begin{equation}
\mathbf{C}\,\mathbf{v}_i = \lambda_i\,\mathbf{v}_i.
\end{equation}

Here, $\lambda_i$ is the eigenvalue associated with the $i$-th principal component and measures the variance explained by that component, while $\mathbf{v}_i$ is the corresponding \emph{principal component} or \emph{eigen-spectrum}. Because $\mathbf{C}$ is symmetric and positive semi-definite, its eigenvectors form an orthonormal basis. The time-variable component of the spectra can therefore be expressed as
\begin{equation}
\tilde{S}(t,\lambda) = \sum_{i=1}^{N_{\mathrm{PC}}} a_i(t)\,\mathbf{v}_i(\lambda),
\end{equation}
where the coefficients $a_i(t)$ represent the temporal amplitudes of each principal component.

In \texttt{Tempawral}, we focus primarily on the eigen-spectra $\mathbf{v}_i$, as these directly encode the dominant wavelength-dependent variability patterns in the time-series data. Only components with eigenvalues significantly above the noise level are retained for subsequent analysis. A key advantage of PCA is its ability to compress hundreds or thousands of spectral snapshots into a compact set of physically interpretable variability modes. For example, the first principal component may trace large-scale brightness changes driven by evolving cloud coverage, while the second may capture spectra variations associated with temperature or chemical modulations. 

In this work, we adopt the first two eigen-spectra ($N_{\mathrm{PC}} = 2$), which typically capture the majority of spectral variability. Figure~\ref{fig:explained_ratio} shows the explained variance for each component in the cases of JWST SIMP~0136 data (Section~\ref{sec:data}) and our canonical simulated model data (Section~\ref{subsec:Eigen}), the first two components explain $\ge 81\%$ and $\ge 99\%$ of the variance, respectively. However, we note the optimal number of components used in \texttt{Tempawral} should be evaluated on a case-by-case basis, balancing the cumulative variance captured with the data quality and computational cost in the retrieval.

\begin{figure*}
	\includegraphics[width=2.1\columnwidth]{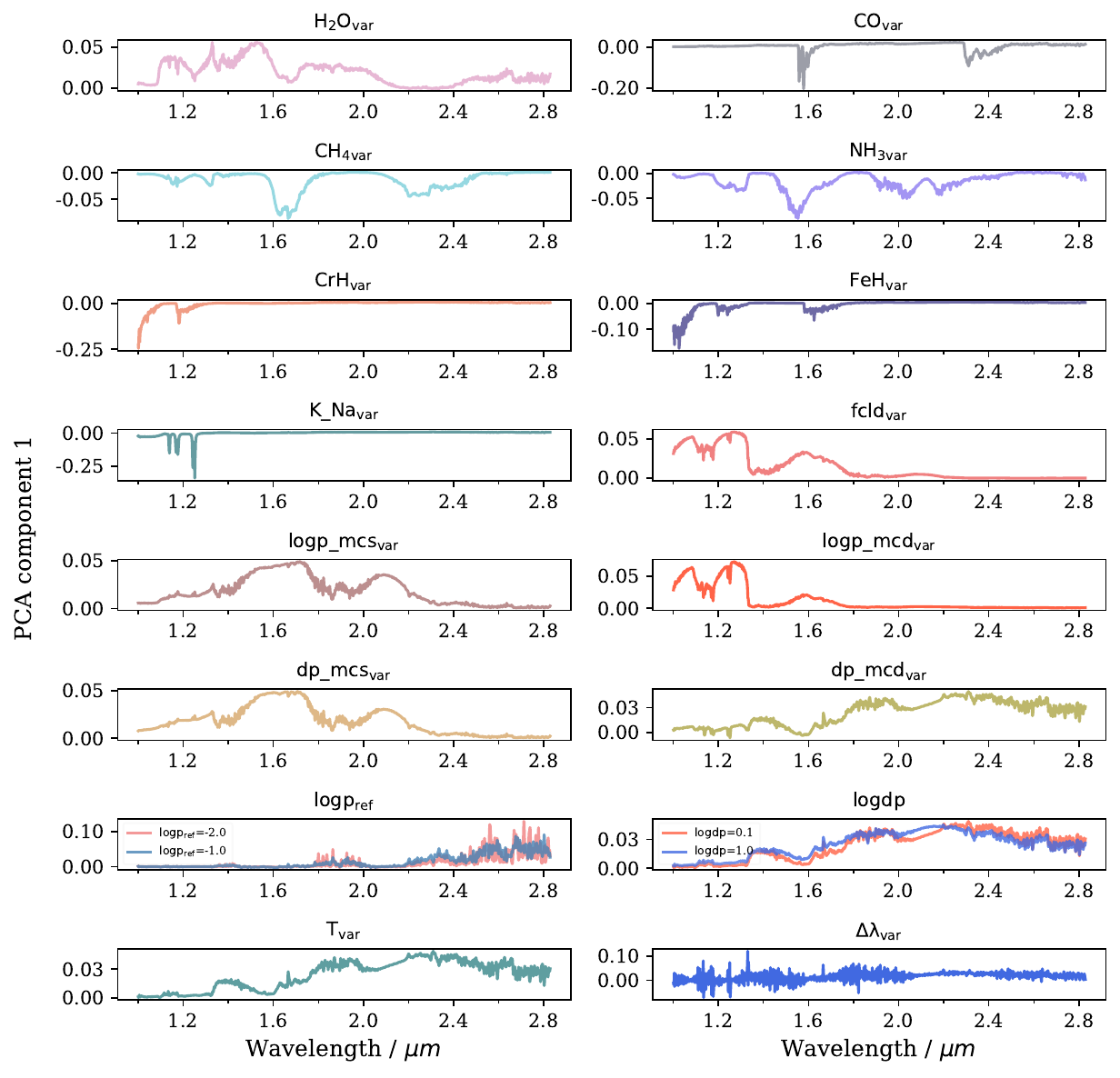}
    \caption{Eigen-spectra sensitivity test on the variance of each model parameters. Starting from a baseline set of atmospheric parameters, time-series spectra are simulated by allowing cloud properties, chemical abundances, and the temperature structure to vary sinusoidally over a full rotation. Each time-variable parameter is perturbed by 10\% from its reference value. In each subplot, we only assign variance on the selected parameter while fixing all others to their reference values, and use this to generate synthetic time-series spectra. The first principal component is then extracted from the resulting time-series. While the parameters $\log p_{\rm ref}$ and $\log dp$, which define the location and extent of temperature perturbations, are modeled as single values and are not treated as time-variable parameters with associated amplitudes and phase shifts.}
    \label{fig:sens}
\end{figure*}

\subsection{Statistical Inference on Eigen-Spectra}

The final step of \texttt{Tempawral} is to infer the perturbation amplitudes and phase shifts of the time-variable atmospheric parameters that best reproduce the observed eigen-spectra. By working in the space of eigen-spectra rather than raw fluxes, the retrieval focuses on the dominant modes of spectral variability, improving computational efficiency and reducing sensitivity to noise and secondary fluctuations.

The retrieval module compares the observed ($\mathbf{v}_{\mathrm{data}}$) and modeled ($\mathbf{v}_{\mathrm{model}}$) eigen-spectra using a Gaussian likelihood function:

\begin{equation}
\ln \mathcal{L}_1 = -\frac{1}{2} \sum_{i=1}^{n} 
\frac{\left(\mathbf{v}_{\mathrm{data},\lambda,i} - \mathbf{v}_{\mathrm{model},\lambda,i}\right)^2}{2\sigma(\lambda_i)^2}
+ \ln \left[ 2\pi \sigma(\lambda_i)^2 \right],
\label{equ:l1}
\end{equation}
where $\sigma(\lambda_i)$ is the propagated uncertainty on the eigen-spectra derived from the original observational errors.

Because $\mathcal{L}_1$ is sensitive only to variability patterns and not absolute flux amplitudes, we include an additional likelihood term, $\mathcal{L}_2$, which compares the peak-to-peak variability amplitude at each wavelength in the original flux domain:

\begin{equation}
\ln \mathcal{L}_2 = -\frac{1}{2} \sum_{i=1}^{n}
\frac{\left(\mathcal{F}_{\mathrm{diff},\lambda_i} - f_{\mathrm{diff},\lambda_i}\right)^2}{2\,\mathrm{err}(\lambda_i)^2}
+ \ln \left[ 2\pi\,\mathrm{err}(\lambda_i)^2 \right].
\label{equ:l2}
\end{equation}

Here, $\mathcal{F}_{\mathrm{diff},\lambda}$ and $f_{\mathrm{diff},\lambda}$ are the observed and modeled peak-to-peak spectral amplitudes, and $\mathrm{err}(\lambda_i)$ is the associated uncertainty. The total log-likelihood is then $\ln \mathcal{L} = \ln \mathcal{L}_1 + \ln \mathcal{L}_2$. We explore the posterior distributions of all time-variable amplitudes and phase shifts using nested sampling (\texttt{PyMultiNest}; \citealt{feroz2009multinest,buchner2014x}). The resulting posteriors quantify the atmospheric variability patterns that best reproduce the observed spectral dynamics.

\subsection{\texttt{Tempawral} Workflow}

The overall procedure outlined is as follows:

\begin{enumerate}
\item Step 1: Reference Values from Baseline Retrieval

From the time-series spectroscopic data, we obtain the baseline atmospheric state by retrieving on the time-averaged spectrum. The maximum-likelihood parameter estimates from this retrieval define the reference set for all model parameters.

\item Step 2: Iterative Time-resolved Retrieval

The following loop is repeated until the PCA components converge. Convergence is assessed on the PCA components rather than on the flux spectra themselves:

\begin{enumerate}
\item Generate a time-dependent model by adding sinusoidal perturbations to the baseline parameters. Each time-variable parameter evolves around its reference value, representing periodic modulation caused by rotating surface features. The resulting time-evolution matrix $\mathbf{L}$ of all model parameters is passed to the time-series forward model to generate the simulated time-series spectra $\mathbf{S}$.

\item Compute the theoretical PCA components ($\mathbf{v}_{\mathrm{model}}$) from the simulated spectra $\mathbf{S}$ using PCA.

\item The retrieval module is then used to compare the theoretical PCA components ($\mathbf{v}_{\mathrm{model}}$) with the observed PCA components ($\mathbf{v}_{\mathrm{data}}$) and infer posterior distributions for the perturbation amplitudes and phase shifts of each time-variable parameters. 

\end{enumerate}
\end{enumerate}

\begin{table}
\begin{threeparttable}
\centering
\caption{Input Time-variable Variance and Phase-shift in Canonical Retrieval }
\begin{tabular}{llrr}
\hline
\hline
\textbf{Category} & \textbf{Parameter} & \textbf{Variance}  & \textbf{Phase-shift} \\
\hline
\multirow{7}{*}{Gas} 
  & $\rm H_{2}O_{var}$        & 0.10 &  0\\
  & $\rm CO_{var}$            & 0.10  &  0\\
  & $\rm CH_{4var}$        & 0.05 &  0 \\
  & $\rm NH_{3var}$        & 0.05  &  0\\
  & $\rm CrH_{var}$           & 0.05  &  0\\
  & $\rm FeH_{var}$           & 0.05  &  0\\
  & $\rm K\_Na_{var}$         & 0.05  &  0\\
\hline
\multirow{1}{*}{Refinement parameter} 
  & $\Delta\lambda_{\rm var}$    & 0.001  &  0\\
\hline
\multirow{3}{*}{Temperature perturbation} 
  & $\log p_{\rm ref}$\tnote{a}         & -0.10  &  \# \\
  & $\log dp$\tnote{a} & 0.10  &   \# \\
  & $T_{\rm var}$              & 200  &  0\\
\hline
\multirow{5}{*}{Cloud} 
  & $f_{\rm cld\_var}$     & 0.05  &  0\\
  & $\log p_{\rm mcs\_var}$ & 0.10  &  0\\
  & $dp_{\rm mcs\_var}$ & 0.10  &  0\\
  & $\log p_{\rm mcd\_var}$ & 0.10  &  0\\
  & $dp_{\rm mcd\_var}$ & 0.10  &  0\\
\hline
\end{tabular}

\begin{tablenotes}
\footnotesize
\item[a] Reference pressure $\log p_{\rm ref}$ and pressure width $\log dp$ are retrieved as single value (not as time-distribution) in the time-resolved retrieval.
\end{tablenotes}

\label{tab2}
\end{threeparttable}
\end{table}

\subsection{Demonstration of Eigen-spectra Sensitivity on Each Model Parameter Variance}

To investigate the sensitivity of the eigen-spectra to variations in individual model parameters, we conduct a parameter sensitivity analysis. We exclude bulk properties such as surface gravity, radius (fixed as ${R^{2}/D^{2}}$ in our case), and Mie-scattering cloud particle size from rotation-induced variability. The baseline $P$-$T$ profile is computed using five atmospheric parameters, while a temperature variance ($T_{\rm var}$) is introduced at a specific reference pressure ($\log p_{\rm ref}$) with a perturbation width of $\log dp$.

In total, 16 parameters are considered: seven gas species, five cloud-related parameters, three temperature perturbation parameters, and a wavelength shift parameter ($\Delta \lambda_{\rm var}$). This wavelength shift models systematics between individual frames due to instrumental effects or time-variable reductions in the observational pipeline. Each time-variable parameter is assigned a sinusoidal variation with a 10\% amplitude around its baseline value from Table~\ref{tab1}. In each experiment, we only assign variance on a single parameter while fixing all others to their reference values, and use this to generate synthetic time-series spectra. We then apply PCA to each synthetic dataset and extract the first principle component in each case. Notably, $\log p_{\rm ref}$ and $\log \Delta p$, which define the vertical location and extent of the temperature perturbation, are not treated as time-variable parameters with associated amplitudes or phase shifts. We explore their effect separately by varying their values and analyzing their impact on the resultant eigen-spectrum.

\begin{figure}
	\includegraphics[width=1 \columnwidth]{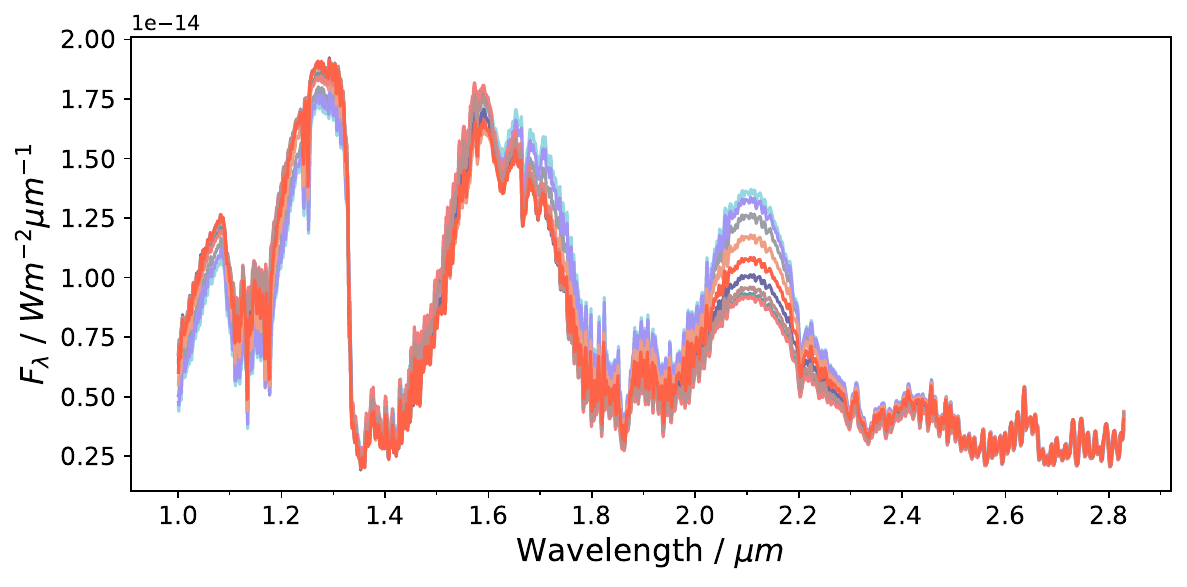}
    \caption{Simulated light curves for a SIMP\,0136-like brown dwarf using our time series forward model. The retrieved median values for each model parameter, as listed in Table~\ref{tab1} from real SIMP\,0136 time-averaged data, are used as the reference values. The variance for each model parameter follows Table~\ref{tab2}, with phase shifts set to zero for all time variables.}
    \label{fig:sim_times_series_spec}
\end{figure}

\begin{figure*}
	\includegraphics[width=2.1\columnwidth]{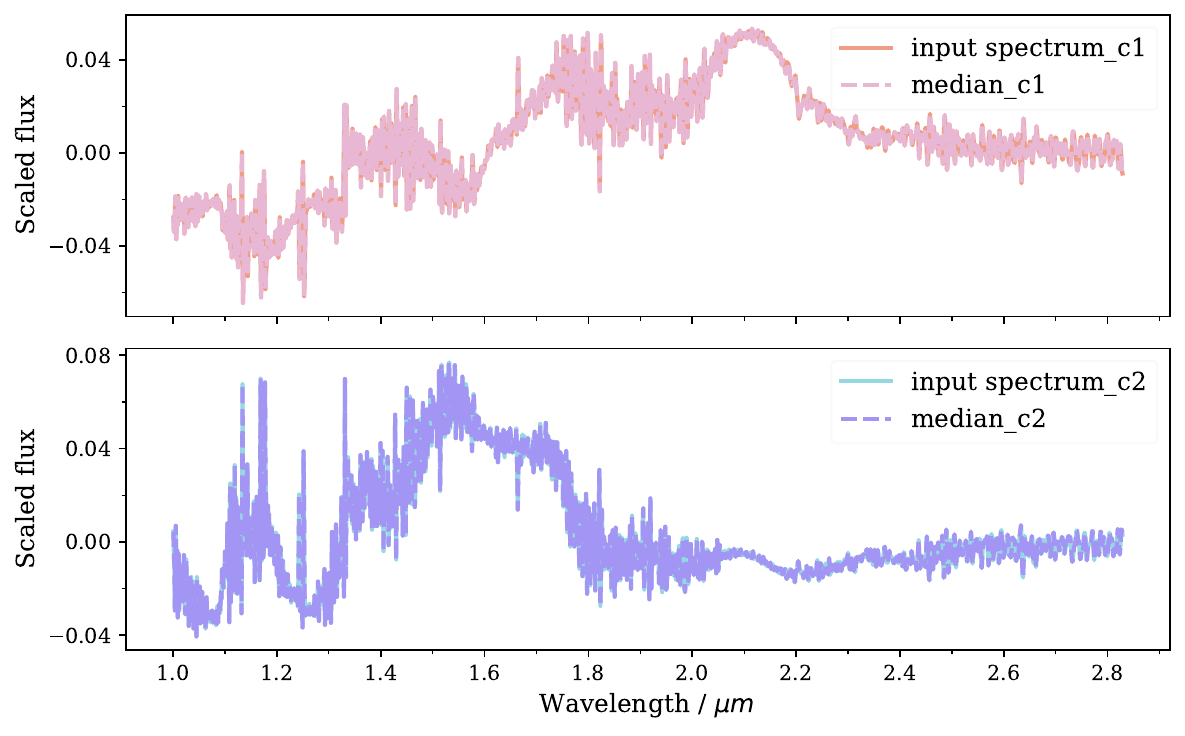}
    \caption{Validation of the \texttt{Tempawral} retrieval framework on the simulated eigen spectrum of a SIMP\,0136-like brown dwarf. The retrieved median model (pink and purple lines for the first and second PCA components, respectively) is compared to the simulated eigen-spectra (yellow and cyan lines for the first and second PCA components, respectively) for the best-fit model.}
    \label{fig:best-fit_sim_eigen}
\end{figure*}

Figure~\ref{fig:sens} shows the eigen-spectra obtained by varying each parameter individually while holding others fixed at their baseline values. We find that each parameter induces a characteristic eigen-spectrum. Variations in gas abundances generally produce eigen-spectra with distinct absorption features within molecular bands, for example $\rm H_{2}O$ at 1.4 and 1.7\,$\mu$m, CO at around 1.6\,$\mu$m, CH$_4$ near 1.6 and 2.2\,$\mu$m, FeH at 1.0 and 1.6\,$\mu$m, and alkali lines centered around 1.0\,$\mu$m. In contrast, cloud-related parameters modulate broader, continuum-like components of the eigen-spectrum. Notably, changes in the cloud base pressure ($\log p_{\rm mcs\_var}$) produce a spectral response similar to that from cloud thickness variations ($dp_{\rm mcs\_var}$), which is reasonable as changes on both two parameter may lead to similar slab-cloud cloud distribution. Likewise, temperature variance introduces broadband effects, whose strength and shape depend sensitively on both the amplitude of the perturbation and its vertical placement ($\log p_{\rm ref}$). These continuum-like components reflect the possible degeneracy in eigen-spectra retrievals. Importantly, we find that even small systematic wavelength shifts ($\Delta \lambda_{\rm var}$) produce continuous, low-level variations in the eigen-spectrum. These broadband, subtle patterns may manifest as systematic noise in real observational data and must be accounted for in retrievals.

These simulations demonstrate the sensitivity of the eigen-spectra to variations in individual model parameters: each parameter modulates the spectral variability in a distinct way, leading to characteristic eigen-spectra. Potential degeneracies among different physical drivers of variability have also been highlighted, for example changes in the cloud vertical distribution parameters can produce similar eigen-spectra.


\begin{figure*}
	\includegraphics[width=2.1\columnwidth]{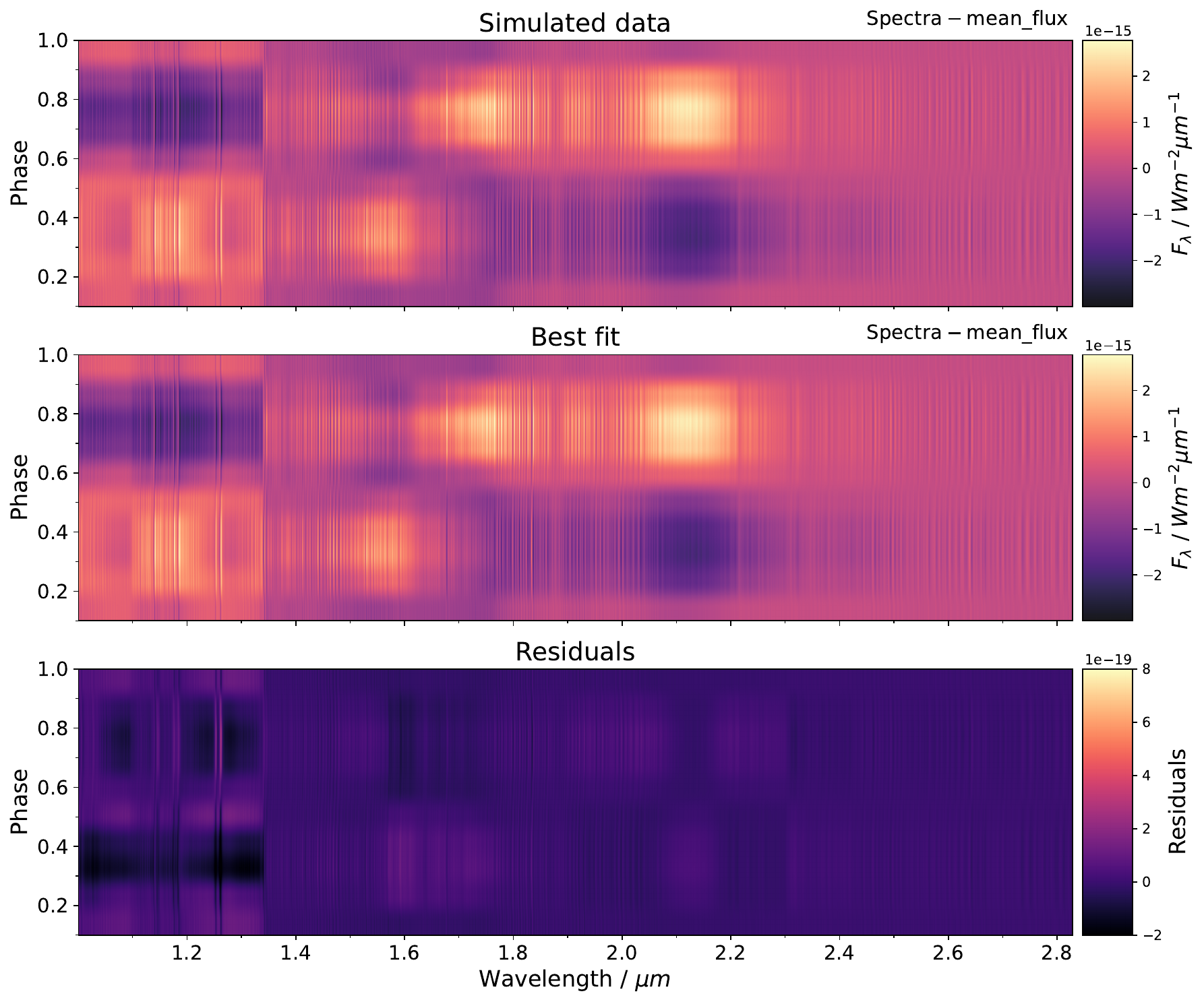}
    \caption{\textbf{Top}: Simulated time-series spectroscopy observation of a SIMP\,0136-like brown dwarf for our Scenario 1. The difference between the each spectrum and time-averaged mean flux is shown at each phase to highlight the variability patterns. \textbf{Center}: the corresponding best-fit solution from the \texttt{Tempawral} atmospheric retrieval. \textbf{Bottom}: residuals between the simulated observations and the best-fit solution.}
    \label{fig:tspec_fit_sim}
\end{figure*}

\section{Validation of Eigen-spectra Retrieval Method on Simulated Data} 
\label{sec:demo}

In this section, we validate \texttt{Tempawral} on simulated time-series spectra for a SIMP\,0136-like brown dwarf. We examine potential parameter degeneracies and explore how well \texttt{Tempawral} can disentangle different physical drivers of variability.

\begin{figure*}
	\includegraphics[width=2.1\columnwidth]{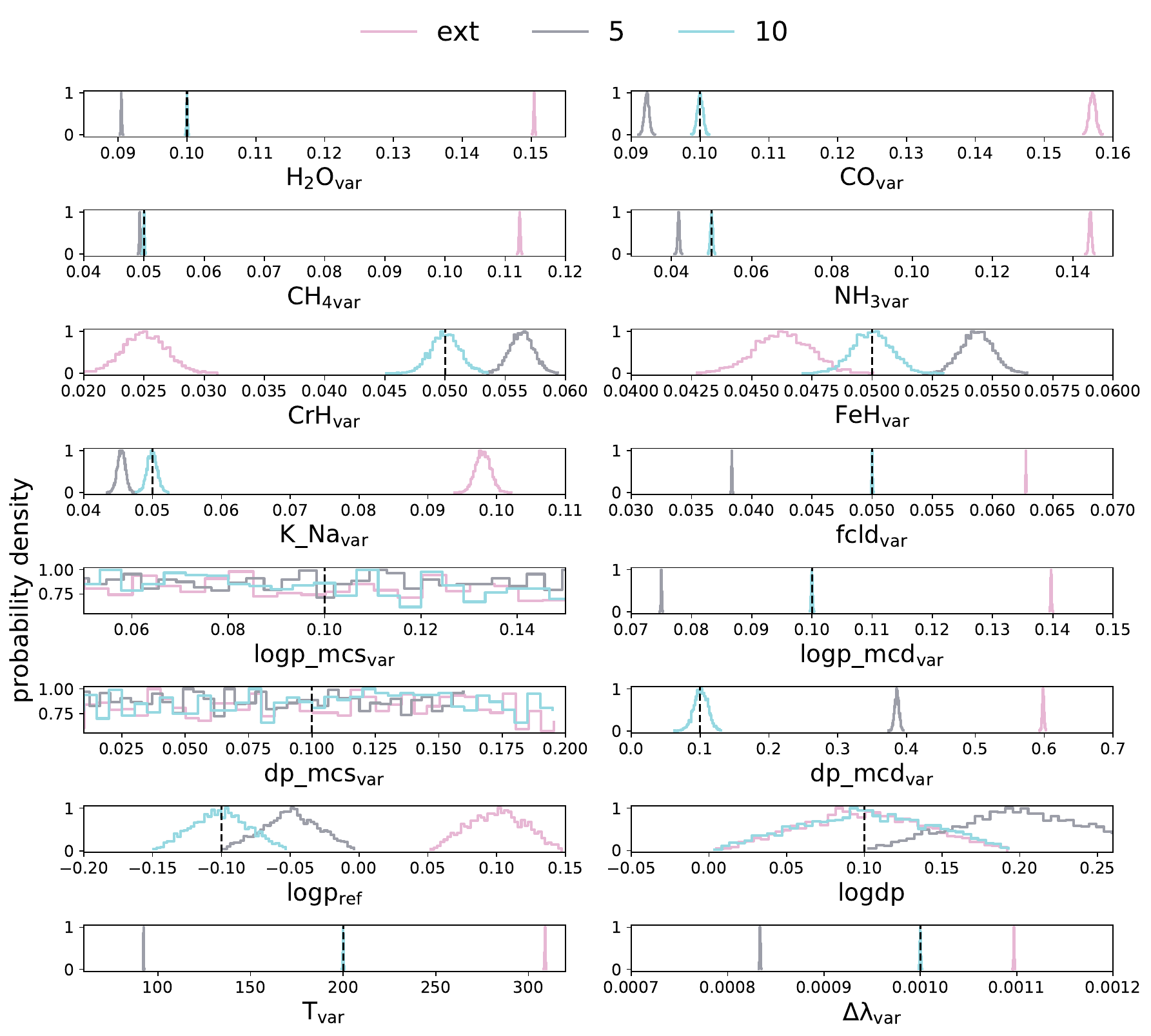}
    \caption{Summary of the marginalized posterior probability distributions as a function of the number of sampling points in the time-series distribution. The black dashed lines indicate the true values (see Table~\ref{tab2}). We test three sampling strategies: \textit{ext}, which samples only the two extrema of the sinusoidal distribution, and uniform sampling of sinusoidal distributions with 5 or 10 points.}
    \label{fig:samplebin}
\end{figure*}

\subsection{Simulated Time-series Observations}

In this simulated experiment, we do not attempt to replicate the exact variability observed in SIMP\,0136. Instead, we use the atmospheric properties retrieved from the real time-averaged spectrum (Table~\ref{tab1}) as a baseline and manually assign variances to key time-variable parameters in order to generate synthetic variability with known ground-truth values.

All gas species included in the retrieval are assumed to vary sinusoidally over the rotational phase. $\rm H_{2}O$ and $\rm CO$ are assigned relatively large abundance variances of 0.1\,dex, while the other gases vary with amplitudes of 0.05\,dex. To simulate inhomogeneous cloud cover, the cloud fraction is treated as time-variable, along with the optical thickness and vertical distribution parameters for both the Mg$_{2}$SiO$_{4}$ slab cloud and the deep $\rm Fe$ deck cloud. The cloud scattering properties and particle sizes are held fixed to reduce degeneracies. The thermal structure is perturbed using the three-parameter temperature modulation scheme described in Section~\ref{subsec:timeV_params}. A 200\,K perturbation is centered at a reference pressure of 100\,mbar ($ \log p_{\rm ref} = -1.0 $) with a pressure range of $\log dp_{\rm var}=0.1$. Additionally, a variance in the spectral wavelength calibration ($\Delta \lambda_{\rm var}=10^{-3}$) is included to account for instrumental systematics. Table~\ref{tab2} summarizes the full set of assigned variances. In this canonical setup, all phase shifts are set to zero for simplicity.

To simulate the time-series spectra, we sample 10 evenly spaced rotational phases from each of the sinusoidal parameter distributions. Figure~\ref{fig:sim_times_series_spec} shows the resulting synthetic time-series spectra. Significant flux variability is evident across the entire spectral range from 1.0 to 2.8 $\rm \mu m$, consistent with the combined influence of all time-variable parameters listed in Table~\ref{tab2}. An average signal-to-noise ratio (S/N) of 200 is assumed for the canonical model. The impact of varying S/N will be explored in subsequent tests. A Monte Carlo simulation is used to propagate noise in the time-series data into the eigen-spectra. We generate 10000 noisy realizations of the time-series spectra by drawing random flux uncertainty from a uniform distribution over the range $[-1\sigma, 1\sigma]$. PCA is performed independently on each realization to extract the corresponding eigen-spectra. From the resulting distribution of eigen-spectra, we compute the median as the central estimate and use the 1$\sigma$ confidence interval to characterize the propagated uncertainty due to observational noise.

\begin{figure*}
	\includegraphics[width=2.1\columnwidth]{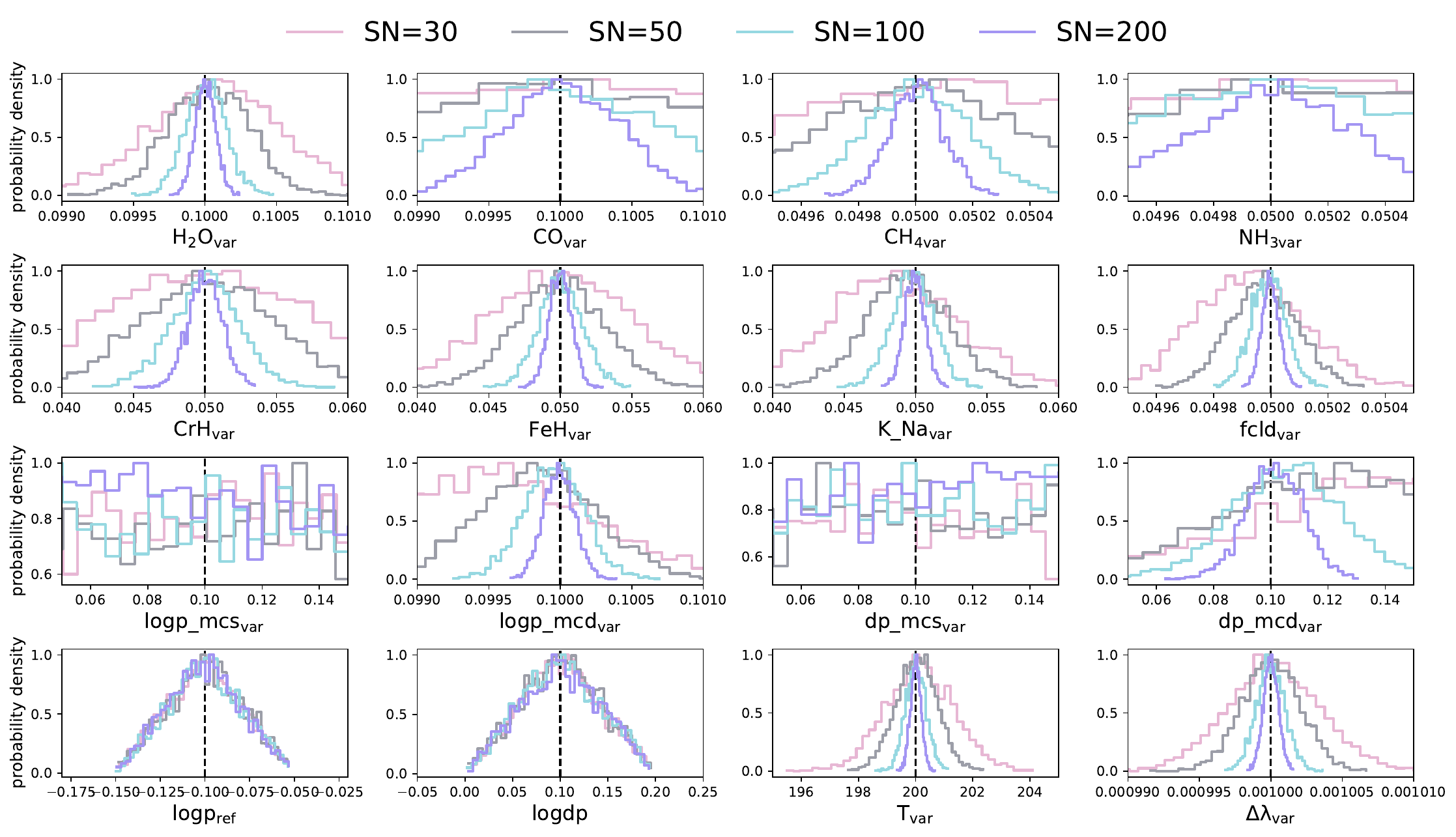}
    \caption{Summary of the marginalized posterior probability distributions as a function of spectral precision in the time-series data, with signal-to-noise ratios ranging from 30 to 200. The black dashed lines indicate the true values (see Table~\ref{tab2}). As expected, parameter constraints degrade as spectral uncertainties increase.}
    \label{fig:SNs}
\end{figure*}



\subsection{Eigen-spectra Retrieval Model Setup}
\label{subsec:Eigen}

\textbf{Scenario 1: Canonical Retrieval}

In this baseline scenario, we perform an eigen-spectra retrieval on simulated time-series spectra of a SIMP\,0136-like brown dwarf. One aim of this scenario is to explore the optimal number of time-sampling bins required to robustly constrain the time-variable parameters. We also intend to explore the influence of data quality on retrieval accuracy. The phase shifts of all time-variable parameters are fixed to zero, representing a simplified case where all variability components are assumed to be synchronized. In total, this run includes 16 free parameters, encompassing the variances of atmospheric chemical abundances, temperature perturbation, cloud dynamics (cloud fraction and vertical structure), and wavelength calibration shift.

We test three sampling strategies in our canonical model: (1) $ext$, which includes only the two extreme phases (only sample the minimum and maximum value of the sinusoidal distribution); (2) $5$-$bins$ sampling, in which five evenly spaced phases are sampled from each time-variable distribution; and (3) $10$-$bins$ sampling, which samples ten evenly spaced phases, providing increasingly dense coverage of the sinusoidal distributions. While increasing the number of sampling points can, in principle, better capture the full spectral variability, doing so comes at a computational cost since generating the simulated time-series spectra must be repeated in every iteration of the Bayesian inference process.
This test helps evaluate the trade-off between retrieval performance and computational efficiency.

JWST time-series observations of brown dwarfs can achieve signal-to-noise ratios of 100–200 in the near-infrared within a small fraction of the rotation period of many nearby BDs, offering high-quality data for studying atmospheric variability \citep{biller2024jwst,chen2025jwst,mccarthy2025jwst}. To evaluate whether these data quality is sufficient for constraining the origins of variability in brown dwarf atmospheres, we perform eigen-spectra retrievals on the SIMP\,0136-like brown dwarf simulation across a range of spectral noise levels. We simulate time-series spectra with varying signal-to-noise ratios (S/N = 30, 50, 100, and 200). Gaussian noise is injected into the spectra and propagated to the eigen-spectra through Monte Carlo simulations. By applying the eigen-spectra retrieval to these datasets, we assess how data quality affects \texttt{Tempawral}’s ability to reliably infer the drivers of variability.

\begin{figure*}
	\includegraphics[width=2.1\columnwidth]{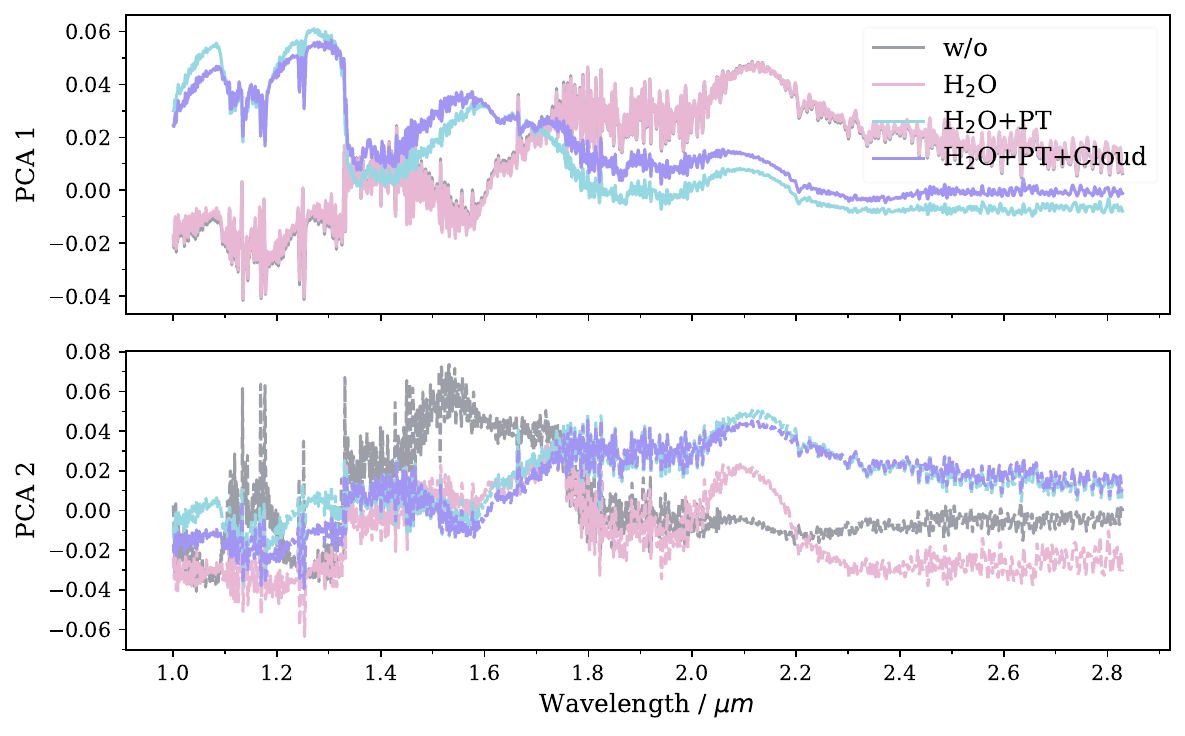}
    \caption{Influence of phase shifts of time-variables on the first two eigen-spectra. The gray curves show the eigen-spectra from the canonical model with zero phase shift for all time-variables ($\rm H_{2}O$, $T_{\rm var}$, and $f_{\rm cld}$). Colored curves represent models with increasing complexity: pink shows the effect of introducing a $\pi/4$ phase shift in $\rm H_{2}O$ alone; cyan includes both $\rm H_{2}O$ and a $\pi/3$ phase shift in temperature; and purple adds a $\pi/2$ shift in cloud fraction as well. The inclusion of phase shift for each time-variable parameter introduces new spectral features in the eigen-spectra, while in the meantime, complicate the retrieval as the introduced possible degeneracies between the variances and phase shifts of time-variables.}
    \label{fig:phaseshift_demo}
\end{figure*}

\textbf{Scenario 2: Phase-Shifted Retrieval}

In contrast to Scenario 1, this case introduces phase shifts among different time-variable parameters to mimic asynchronous variability patterns that may arise from heterogeneous atmospheric structures. For example, temperature perturbations, molecular abundance fluctuations, and patchy cloud distributions may not co-rotate in phase, reflecting complex longitudinal and vertical inhomogeneities.

While a fully general model would assign an independent phase-shift parameter to each time-variable listed in Table~\ref{tab2}, we simplify the analysis by applying a representative phase shift to one key variable from each category:

\begin{itemize}
  \item $\mathrm{H_2O}$ (chemistry) $\rightarrow$ phase shift of $\pi/4$,
  \item $T_{\mathrm{var}}$ (temperature) $\rightarrow$ phase shift of $\pi/3$,
  \item $f_{\mathrm{cld}}$ (cloud) $\rightarrow$ phase shift of $\pi/2$.
\end{itemize}

These are implemented as three additional free parameters: $\rm H_{2}O_{shift}$, $T_{\rm shift}$, $f_{\rm cld\_shift}$, resulting in a total of 19 free parameters for this scenario. This setup allows us to explore the robustness of the retrieval in the presence of more realistic and potentially degenerate phase offsets.

For both scenarios, we employ \texttt{PyMultiNest}, a nested sampling implementation \citep{feroz2009multinest,buchner2014x}, to explore the posterior distribution of the model parameters. We adopt non-informative uniform priors across the allowed ranges and initialize the sampler with 500 live points. The $\mathtt{evidence\_tolerance}$ is set to 0.1, meaning the sampling process terminates once the remaining contribution to the evidence from the unexplored prior volume falls below 10\% \citep{buchner2016statistical}.





\begin{figure*}
	\includegraphics[width=2.1\columnwidth]{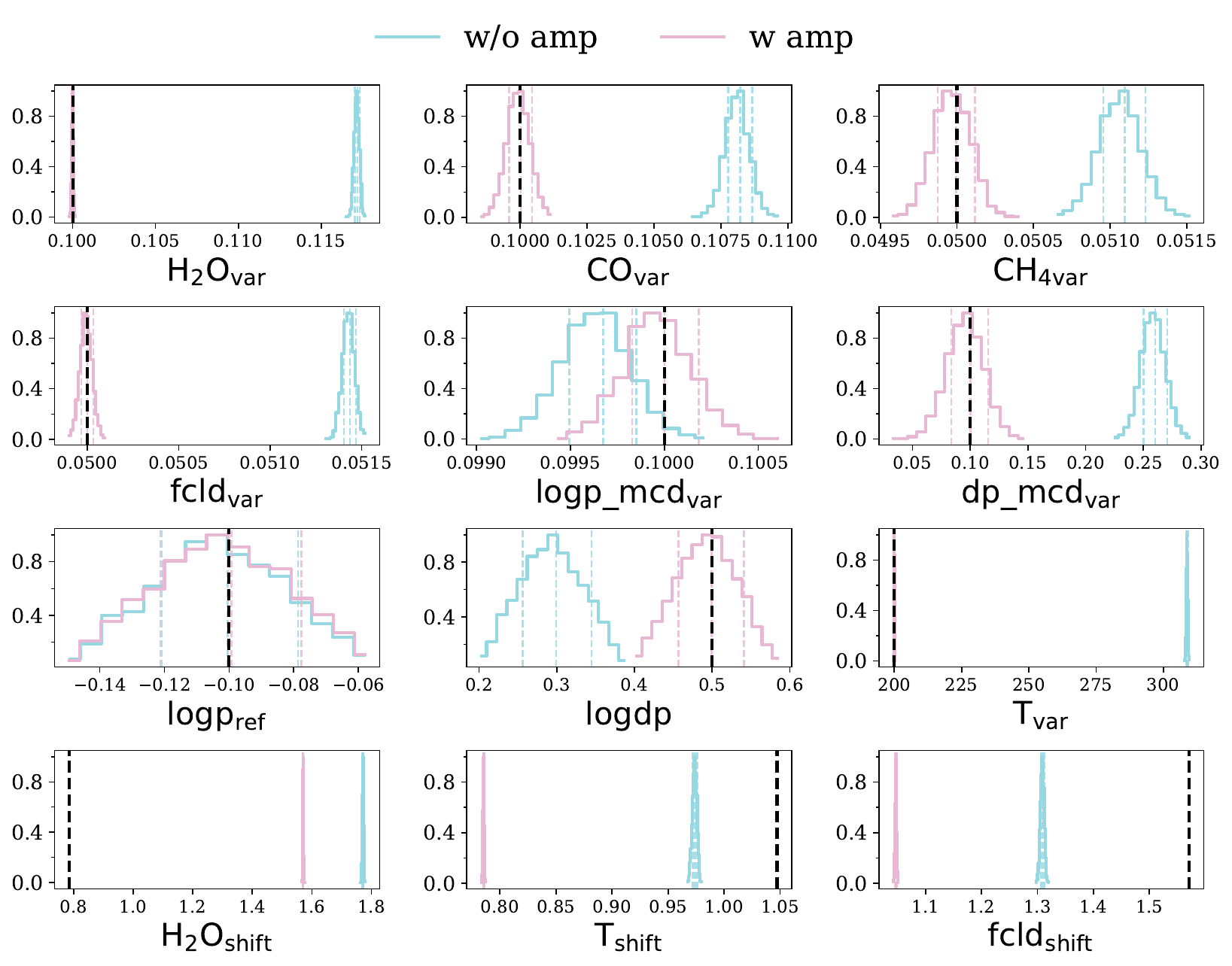}
    \caption{Phase-shifted retrieval: posterior probability distributions for selected time-variables comparing retrievals using only eigen-spectra fitting (solid lines) and those incorporating additional flux amplitude information from the original time-series data (dashed lines). Colored dashed lines mark the 16\%, 50\%, and 84\% quantiles of each distribution, while black dashed lines indicate the true values (see Table~\ref{tab2}). Including amplitude information helps tighten constraints and reduce degeneracies in parameter estimation.}
    \label{fig:amp_plot}
\end{figure*}

\section{Results from Simulated Data Validation Test} \label{sec:result}


Using time-series spectroscopy simulations of a SIMP\,0136-like brown dwarf, we demonstrate the capabilities of \texttt{\texttt{Tempawral}} in deciphering the origin of atmospheric variability.

\begin{figure*}
	\includegraphics[width=2\columnwidth]{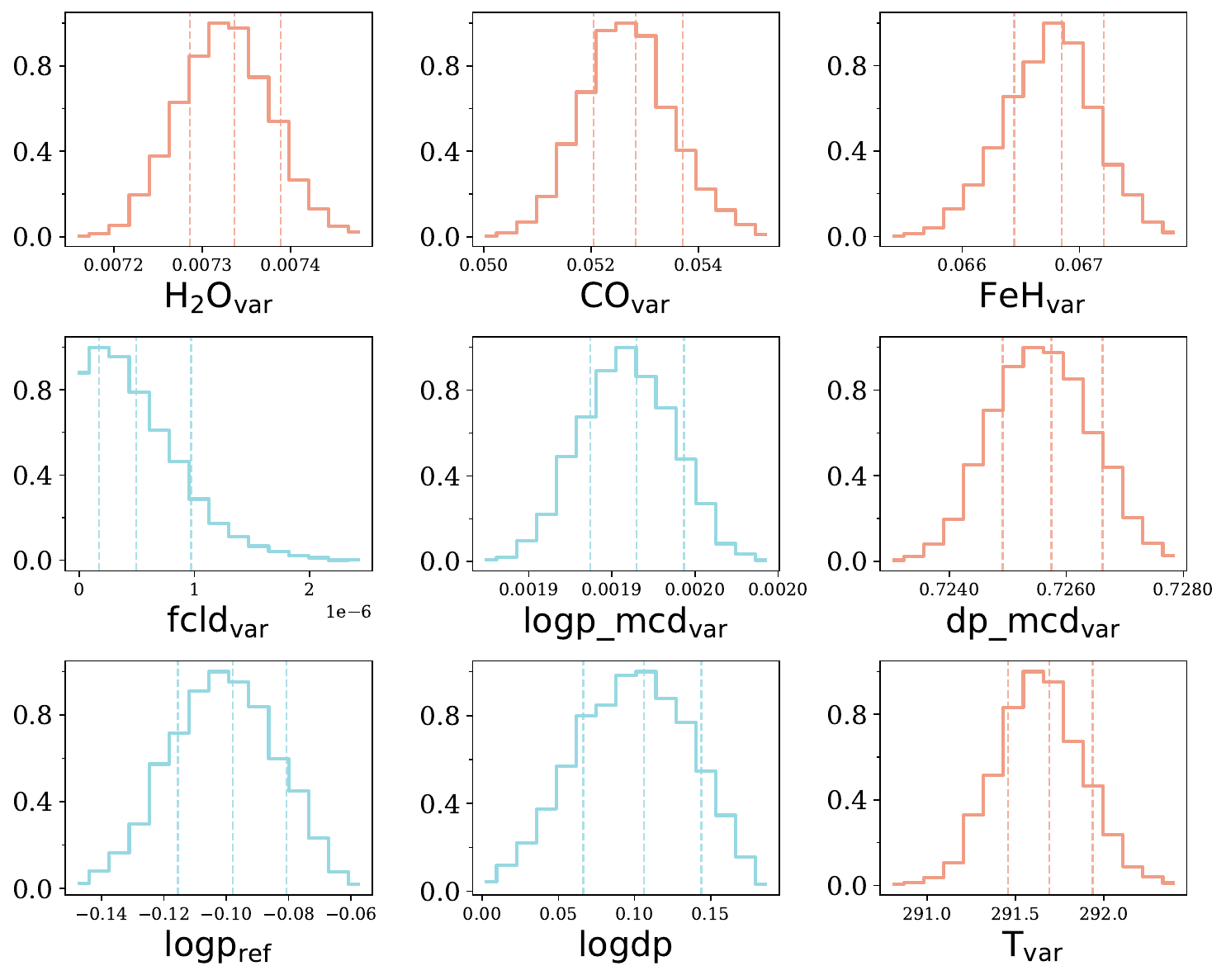}
    \caption{Summary of marginalized posterior probability distributions for the varying time-variables in the JWST near-infrared SIMP\,0136 data. Dashed colored lines indicate the 16\%, 50\%, and 84\% quantiles. Orange lines highlight the primary drivers of variability with significant variance, and cyan lines denote the more quiescent variables, including the cloud fraction ($\rm fcld_{var}$) and the location ($\rm logp_{ref}$,$\rm logdp$ ) where temperature perturbation happens.}
    \label{fig:data_hisplot}
\end{figure*}

\subsection{Scenario 1: Canonical Retrieval} 

We simulate time-series spectra for a SIMP\,0136-like brown dwarf with 10 evenly spaced sampling phases (Figure~\ref{fig:sim_times_series_spec}). The most prominent flux contrast appears around 2.1 $\rm \mu m$, mainly driven by changes in the vertical cloud distribution. 
The wavelengths of fluctuating fluxes correspond to absorption bands of molecules. For instance, absorption features are seen for $\rm H_{2}O$ at 1.4 and 1.7\,$\mu$m, CO around 1.6\,$\mu$m, $\rm CH_{4}$ at 1.6 and 2.2\,$\mu$m, FeH at 1.0 and 1.6\,$\mu$m, and alkali lines near 1.0\,$\mu$m for Na and K. A PCA is applied to the time-series data to extract the eigen-spectra, which then serve as input for the \texttt{Tempawral} retrieval.

Figure~\ref{fig:best-fit_sim_eigen} shows the resultant the eigen-spectra fit. The median eigen-spectra, drawn from 1000 posterior samples, successfully reproduce the shape and features of the data’s first two principal component. Figure~\ref{fig:tspec_fit_sim} presents the best-fit spectro-temporal variability map. To highlight the variability patterns, each spectrum is shown relative to the mean flux. The best-fit variability map reproduces the key features of the data, especially the strong variability around 2.2\,$\mu$m. The residuals are minimal (4 magnitude smaller than the original flux difference) and consistent with white noise, as expected in this idealized case. Figure~\ref{fig:corner_sim} shows the marginalized posterior distributions for all time-variable parameters, including chemistry, cloud structure, temperature perturbations, and wavelength shifts. The black lines represent the ground-truth values. The retrieved distributions are centered around these values. This test demonstrates that \texttt{Tempawral}, when combined with eigen-spectra analysis, can effectively recover the underlying variability patterns from simulated data.

\subsubsection{The Impact of Time-sampling Bins} 



Figure~\ref{fig:samplebin} illustrates how the number of sampling bins affects the accuracy of eigen-spectra retrieval in recovering the variances of each time-variable parameter. When using only the two extreme points (the peak and trough) from each sinusoidal curve, the retrieved marginal distributions deviate significantly from the input values. This is expected, as sampling only the extremes captures limited variability information, which compromises the quality of the PCA and leads to inaccurate eigen-spectra. Increasing the sampling to 5 points improves the results but still yields bias for most of the time-variables shown in gray curve. However, as shown by the cyan-blue curve in Figure~\ref{fig:samplebin}, using 10 evenly spaced samples across the sinusoidal curve enables accurate recovery of all parameter variances. This number strikes a balance between capturing the variability in the data and maintaining computational feasibility, especially given that each retrieval iteration requires simulating 10 snapshot spectra and the total sampling evaluations numbers can reach the order of one million for the full inference process. In this canonical eigen-spectra retrieval using the full time-series dataset, the total computational time is comparable to that of a traditional single-snapshot spectral retrieval, approximately 60 hours on a 128-core computing cluster.

\subsubsection{Influence of Noise Levels}

Figure~\ref{fig:SNs} illustrates the effect of spectral precision of time-series spectra on the posterior distributions of key atmospheric time-variables. Figure~\ref{fig:corner_SNs} shows the marginalized posterior distributions for all time-variable parameters. As expected, lower spectral precision results in broader uncertainties. More notably, for several key parameters, reduced precision not only increases uncertainty but also introduces systematic bias in the posterior distributions due to increased degeneracy. For instance, the vertical location of the deck cloud top and deck cloud thickness can exhibit degeneracy: at lower precision, retrievals favor lower variance in the deck cloud top and a more variable deck cloud vertical distribution. Importantly, even with the lowest-quality data tested (pink and gray histograms), the retrieval still provides meaningful constraints on the chemical variability and temperature perturbations. However, accurate constraints on cloud structure parameters become increasingly difficult under high-noise conditions, due to their intrinsic correlations.

These results underscore the sensitivity of eigen-spectra to subtle atmospheric variability encoded in rotational modulation. However, as also highlighted in Figure~\ref{fig:sens}, certain combinations of atmospheric changes can yield similar eigen-spectra, introducing parameter degeneracies. As shown by the cyan and purple histograms, a minimum S/N of 100 is necessary to constrain all model parameters in our simulated SIMP\,0136-like brown dwarf. This experiment demonstrates that eigen-spectra retrieval is feasible for recent JWST time-series observations with S/N between 100 and 200. However, this S/N requirement depends on the variability amplitude of the brown dwarf. Higher data quality is needed for objects with low variability, such as LSRJ~1835 \citep{reid2003meeting}, which exhibits only $\sim1\%$ near-infrared variability.

\subsection{Scenario 2: Phase-Shifted Retrieval}

A similar retrieval exercise is conducted, now incorporating a more complex atmospheric configuration in which multiple asynchronous surface features coexist. Unlike Scenario 1, where all time-variable phase shifts were set to zero, Scenario 2 introduces phase offsets to selected time-varying parameters. Figure~\ref{fig:phaseshift_demo} illustrates how the eigen-spectra respond to gradual inclusion of phase shifts in the $\rm H_{2}O$ abundance, temperature perturbation, and cloud fraction. The grey curves represent the canonical case with no phase shifts. The pink, purple, and blue curves illustrate progressively more complex models, beginning with a single phase-shifted parameter ($\rm H_{2}O_{shift}$) and extending to models where the three main variables of $\rm H_{2}O$, $f_{\rm cld}$, and $T$, have phase shifts of $\pi/4$, $\pi/3$, and $\pi/2$, respectively.

Importantly, introducing phase shifts alters the overall shape of both the first and second PCA components. For example, in the case where $\rm H_{2}O$ abundances time-distribution changes with a phase shift of $\pi/4$, the first principal component appears similar to the canonical case, but the second component deviates significantly. This highlights the necessity of including phase-shift parameters in eigen-spectra retrieval, as these shifts directly influence the spectral structure. Moreover, fitting both the first and second PCA components simultaneously becomes essential to break potential degeneracies.

Due to this added model complexity  and increased parameter dimensionality compared to canonical model, we incorporate additional flux amplitude information into the likelihood function  and define a composite likelihood ($\log L =\ln L_{1} + \ln L_{2}$), where $\ln L_{1}$ (Equation~\ref{equ:l1}) captures the eigen-spectra fit, and $\ln L_{2}$ ( Equation~\ref{equ:l2}) encodes flux variability amplitude. Figure~\ref{fig:amp_plot} compares two retrievals: one using only eigen-spectra ($\rm lnL_{1}$, labeled "w/o amp") and the other using the combined likelihood ("amp" retrieval). The "amp" retrieval (pink histograms) accurately constrains the variances of chemistry, cloud, and temperature structure parameters, while the canonical model (grey curves) fails to recover the true values. In particular, $\rm H_{2}O_{var}$, $ f_{\rm cld\_var}$, $T_{\rm var}$ are significantly biased when their associated phase shifts are ignored. However, as shown in the lower panel of Figure~\ref{fig:amp_plot}, even though the variance parameters are well constrained in the “amp” retrieval, the recovered phase shifts for the three variables deviate from their true values. This suggests that while the inclusion of variability amplitude improves variance retrieval, it is insufficient for accurately resolving the phase shifts themselves. This limitation may stem from the fact that the retrieval is based on a single rotational coverage of the times-series spectroscopy motioning. Time-series observations covering multiple rotational periods could be helpful to fully disentangle phase-shift effects among time-variable components.


\begin{figure*}
	\includegraphics[width=2.1\columnwidth]{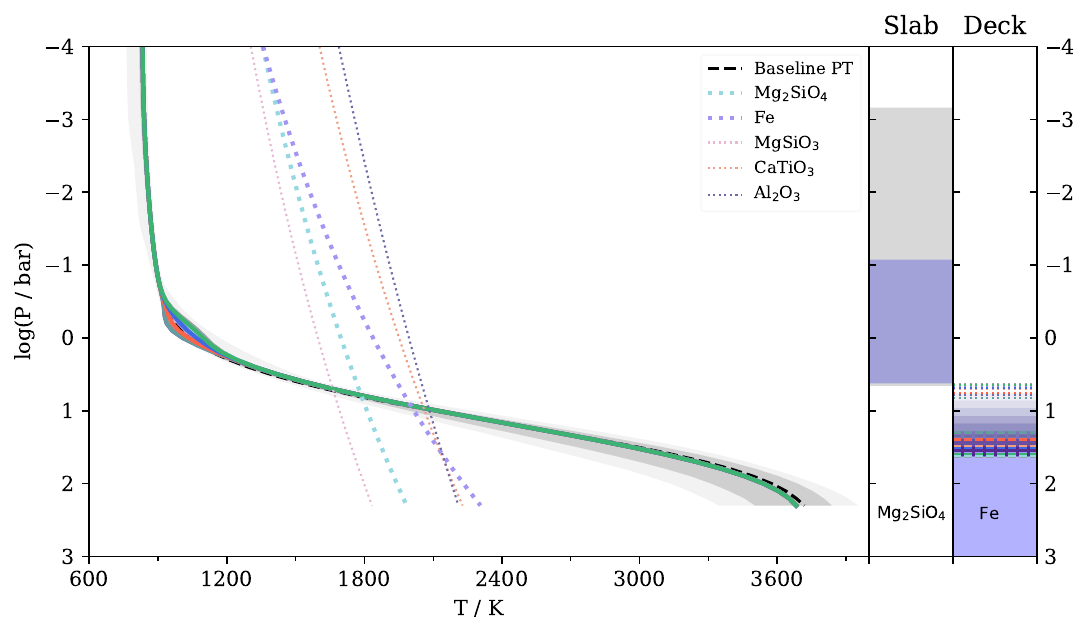}
    \caption{Summary of the evolving temperature structure and iron deck cloud distributions for SIMP\,0136. Colored lines represent the temperature profiles and cloud deck distributions at 10 evenly spaced time steps. Black dashed lines indicate the median pressure-temperature profile from the time-averaged spectrum retrieval, with gray shaded regions showing the $1\sigma$ and $2\sigma$ confidence intervals. Phase-equilibrium condensation curves for plausible cloud species are plotted as coloured dotted lines. In the right panel, clouds pressures are indicated in bars to the right. Purple bar indicates the median cloud location for the forsterite slab and the optically thick extent of the iron deck, with grey shading indicating the $1\sigma$ range. The changing cloud deck distributions are shown as a function of time, where the upper dotted horizontal lines correspond to the $\tau_{\rm cld}= 1/e$ optical depth layers, and the lower dashed horizontal lines mark the $\tau_{\rm cld}= 10$ layers.}
    \label{fig:data_pt_evole}
\end{figure*}

\begin{figure*}
	\includegraphics[width=2.1\columnwidth]{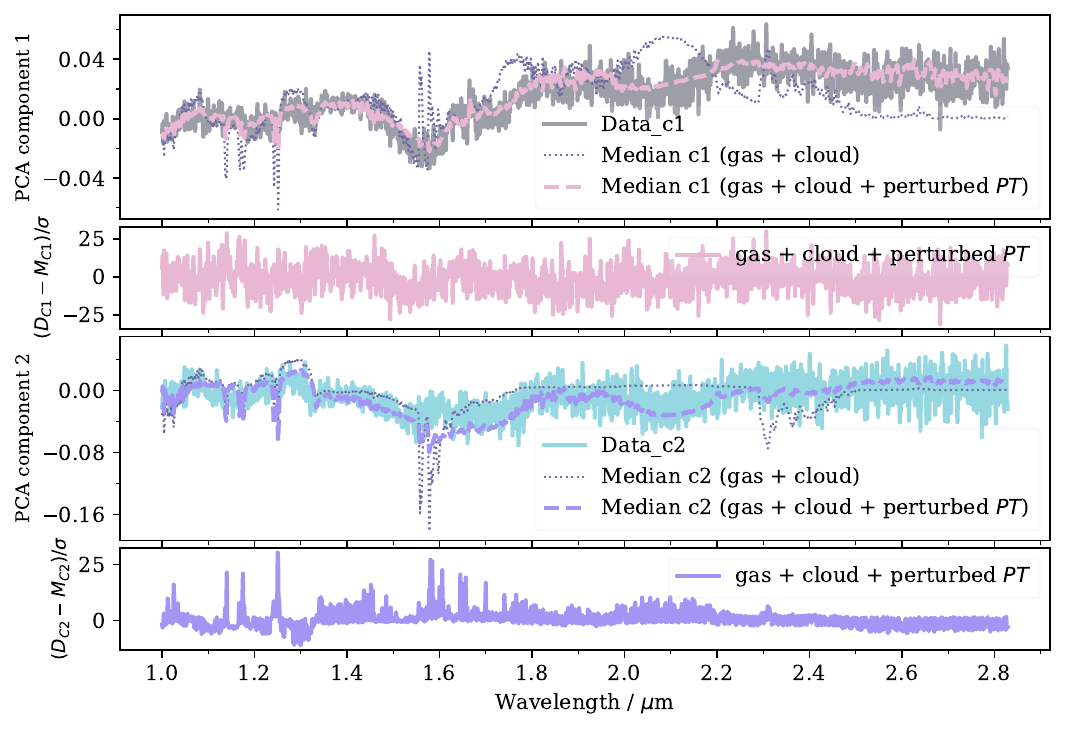}
    \caption{Validation of the \texttt{Tempawral} on the Eigen-spectra from JWST/NIRISS time-series observations of SIMP\,0136. The pink and purple lines denote the first and second PCA components from the retrieved best-fit model (considering time-variables of chemistry, cloud, and temperature perturbation), while the gray and cyan lines represent the corresponding PCA components from the observed Eigen-spectra. A comparison model without temperature perturbation is also shown with dotted lines. The residuals, defined as the difference between the retrieved median eigen spectrum and the data eigen spectrum and normalized by the  data uncertainties, are also shown.}
    \label{fig:data_eigenfit}
\end{figure*}

\begin{figure*}
	\includegraphics[width=2.1\columnwidth]{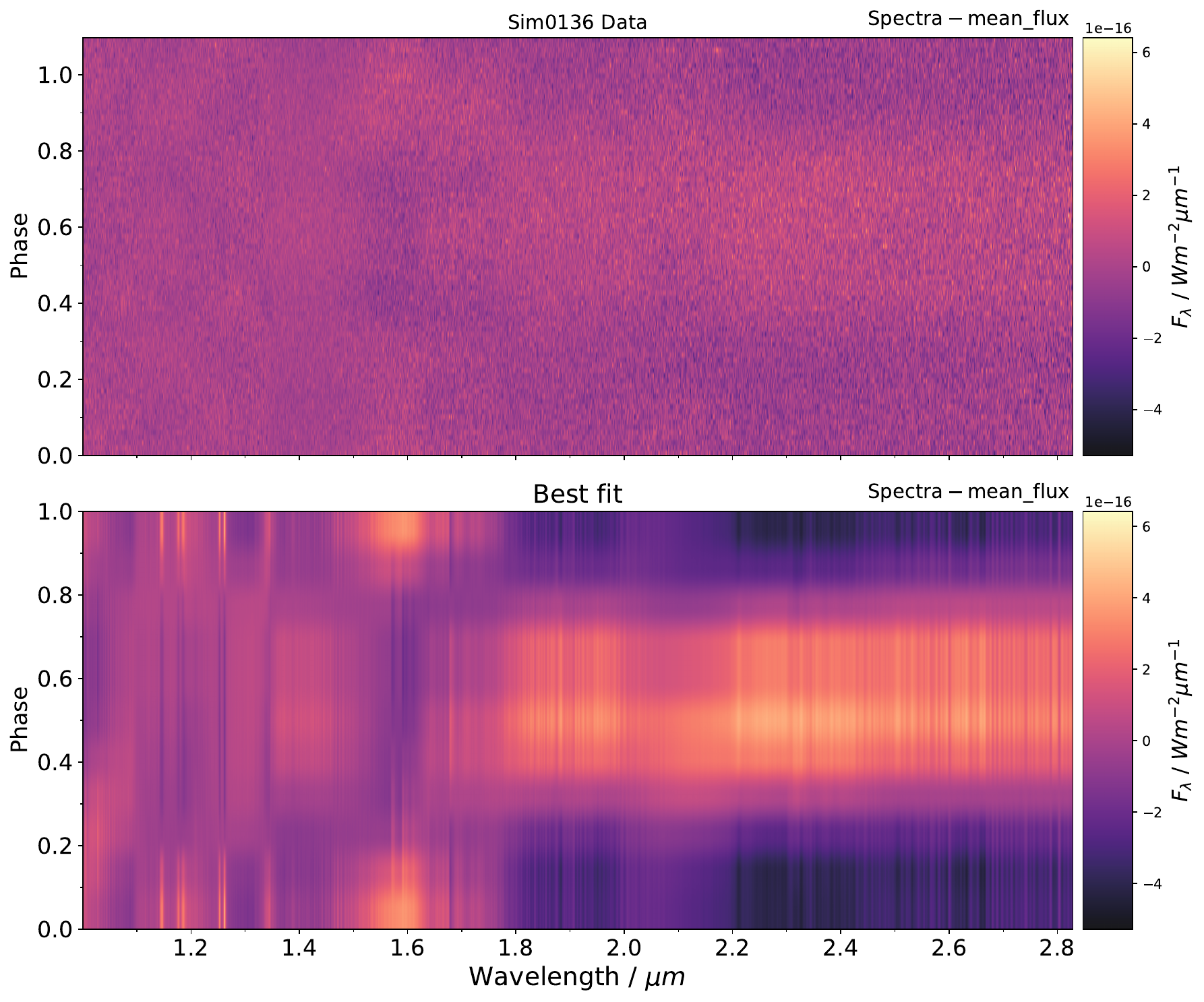}
    \caption{\textbf{Top}: JWST/NIRISS-SOSS time-series spectroscopy observations of SIMP\,0136 with total number of 81 spectra. The differences between each spectrum and the time-averaged mean flux are shown at each phase to highlight variability patterns. \textbf{Bottom}: Corresponding best-fit solution from the \texttt{Tempawral} atmospheric retrieval. The modeled time-series spectra (10 spectra) reproduces the primary variability patterns, notably the strong positive feature near the center and negative features toward the edges within the 1.8 to 2.8 $\rm \mu m$ range. These variability patterns  are less visually apparent in the data due to the presence of observational noise.}
    \label{fig:data_tspec_fit}

\end{figure*}

\section{Application to JWST/NIRISS-SOSS Time-Series Data of SIMP\,0136} 
\label{sec:application}

In this section, we apply \texttt{Tempawral} directly to the real time-series dataset as detailed in Section~\ref{sec:data}, aiming to identify the origin of variability in SIMP\,0136 and to quantitatively constrain the time-variable parameters.
 

\subsection{Baseline Retrieval} 

We use the same baseline retrieval model (Table~\ref{tab1}) retrieved on the median spectrum of time-series data of SIMP\,0136. This baseline model followed the prescription described in \citet{vos2023patchy}, the atmosphere of SIMP\,0136 is best explained by a two-layer cloud structure: a high-altitude forsterite (Mg$_{2}$SiO$_{4}$) cloud overlaying a deeper, optically thick iron (Fe) cloud, alone with the dominant molecular absorptions from H$_{2}$O, CO, CO$_{2}$, CH$_{4}$, NH$_{3}$, CrH, FeH, Na, K.
 
\subsection{Eigen-spectra Retrieval Setup}

Similar to the simulations setup in Section~\ref{subsec:Eigen}, we computed the eigen-spectra using the PCA analysis on the SIMP\,0136  time-series data   and obtain the noise of eigen-spectra by propagating noise of time-series spectra through Monte Carlo simulations. Next, we performed a series of phase-shifted retrievals that considered changes in gas abundances, temperature structure, and cloud properties. 
The cloud scattering properties and particle sizes are held fixed to reduce degeneracies. The variances of chemical abundances follow a log-uniform prior, with $\log(VMR_{\rm var}) \in [0, 1])$. A wavelength shifts ($\Delta \lambda_{\rm var}$) is considered to account for the continuous, low-level variations resulted from the systematic noise in real observational data.

\subsection{Result} 

The best variability model for SIMP\,0136 is comprised of a $\sim$300 K perturbation near 1\,bar, combined with changing abundances of $\rm H_{2}O$, $\rm CO$, $\rm FeH$, and variations in the thickness of the iron deck cloud. Figure~\ref{fig:corner_data} presents the posterior distributions of variances and phase shifts for key parameters, including gas abundances, cloud properties, and temperature. Specifically, the orange histograms in Figure~\ref{fig:data_hisplot} highlight the most prominent variable parameters across chemistry, cloud properties, and temperature. Among chemical species, $\rm FeH$ exhibits the largest variability, with $\rm FeH_{var}=0.0668^{+0.0007}_{-0.0009}$, followed by $\rm CO $ with $\rm CO_{var}= 0.0526^{+0.0016}_{-0.0020}$. In contrast, $\rm H_{2}O$ varies marginally, at a level of just $\rm H_{2}O_{var}=0.007325^{+0.000097}_{-0.000119}$. The corresponding phase shifts for $\rm FeH$  and the $\rm CO $ are $\rm FeH_{shift}=1.58^{+0.02}_{-0.04}$ and $\rm CO_{shift}= 1.73^{+0.03}_{-0.04}$, respectively. As discussed in Section 4.2, including phase-shift parameters introduces additional degeneracies, potentially biasing their estimates. Given that this dataset only spans a single rotational period, we treat the recovered phase-shift values cautiously. Multiple rotational phase observations would be required to robustly constrain these parameters.

The best-fit model includes a patchy forsterite slab cloud overlying a deeper, variable iron deck cloud. We find no evidence for variability in cloud fraction, or in the scattering properties and vertical distribution of the forsterite slab cloud, which is in line with 
the multiple-epoch retrievals of SIMP\,0136 with JWST NIRSpec/PRISM and MIRI/LRS time-series observations (0.8-11 $\rm \mu m$) \citep{nasedkin2025jwst}. Patchy silicate clouds were required to fit the observed spectra, but the scattering properties and vertical distribution of the forsterite slab cloud were not found to systematically vary in the rotation. However, the decay height of the iron deck cloud, $dp_{\rm mcd}$, shows significant variation, with $ dp_{\rm mcd\_var}=0.72^{+0.0016}_{-0.0015}$. This decay height describes how optical depth changes with pressure: $d\tau /dp \propto \exp((P-P_{\rm deck})/\Phi)$, where $P_{\rm deck}$ is the pressure level at which the optical depth at 1 $\rm \mu m$ first exceeds unity, and $\Phi$ is a scale parameter which tracks how the opacity builds up with increasing pressure. Variances in $dp_{\rm mcd}$ indicates substantial changes in how the cloud’s optical depth ramps up and down across atmospheric layers. To visualize this effect, we plot the cloud top ($\tau=1/e$) and base ($\tau=10$) for each of the 10 time steps in the right panel of Figure~\ref{fig:data_pt_evole}. A notable temperature perturbation is recovered, centered at $\log p_{\rm ref}=-0.10^{+0.02}_{-0.01}$, with a pressure range of $\log dp=0.10^{+0.04}_{-0.04}$. and amplitude of $ T_{\rm var}=291.81^{+0.78}_{-0.40}$ K. We overplot the perturbed temperature profiles at 10 rotational phases on top of the baseline $P$-$T$ profile in Figure~\ref{fig:data_pt_evole}, the temperature structure shows a clear bump around 1 bar. Together with the recovered changes in $\rm FeH$ and the $\rm CO$ abundances and the varying iron deck cloud structure, all these changing atmospheric properties contribute to the spectro-temporal variability in the observed data.

Figure~\ref{fig:data_eigenfit} compares the  best-fit eigen-spectra with observed data eigen-spectra. The model reproduces the overall shape of the first and second PCA components. However, it fails to fit the observations between 1.6–1.8\,$\mu$m. This discrepancy likely arises from complex degeneracies the missing model complexity of cloud scattering properties and its particle radii distribution, or a parameter that is varying differently in the two components that a single sinusoidal variation with one phase-shift and variance cannot capture. For example, this band lies in the water abortion band, the poor fit at this band by fitting the two eigen-spectra simultaneously may suggest there could be multiple asynchronous water surface features coexist in the atmosphere, a single set a variance and phase-shift is not sufficient to mimic the complex patterns of global water abundance distributions. Another interesting feature is the presence of subtle, continuous patterns in both PCA components of the data, and also result in the  large fluctuate residuals shown in both two component fitting. We note this may be linked to the random distribution of wavelength shifts between individual spectra. Although we model this shift as a sinusoidal $\Delta \lambda$ (on the order of $10^{-6}$), this may not adequately capture the true stochastic behavior in the observations, as discussed in Figure~\ref{fig:sens}.

In the last, Figure~\ref{fig:data_tspec_fit} compares the variability patterns from the data and the best-fit spectro-temporal flux maps reproduced from the best-fit model. We highlight variability by plotting the difference between each spectrum and the mean flux. The best-fit model reproduces the primary variability patterns well, particularly the strong positive feature in the center and negative features above and below it across the 1.8–2.8 $\rm \mu m$ range. In the 1.0–1.6 $\rm \mu m$ region, finer structures are also consistent between the model and data. However, the 1.6–1.8 $\rm \mu m$ range again shows notable discrepancies as the poor fit of second eigen spectrum between the data and the model. Figure~\ref{fig:lightcurve} further compares the reconstructed light curves across five photometric bands, post-processed from Figure~\ref{fig:data_tspec_fit}. The model shows a consistent variability trend across all bands when compared with the observed light curves.
\section{Discussion}
\label{sec:dis}

\subsection{Comparison with previous work}
Recent studies have examined the atmospheric variability of SIMP\,0136 using broader time-series spectral coverage (0.8–11 $\rm \mu m$), capturing a full rotation cycle with both NIRSpec/PRISM and MIRI/LRS instruments aboard JWST. \citet{mccarthy2025jwst} analyzed this dataset using a light-curve clustering technique and concluded that the pressure-dependent variability is driven by multiple mechanisms of cloud dynamic, localized hot spots, and changes in carbon chemistry. These findings align well with our interpretation: a $\sim$300 K temperature perturbation near 1 bar, combined with time-variable abundances of $\rm H_{2}O$, $\rm CO$, $\rm FeH$, as well as variations in the thickness of the iron deck cloud, contributes to the observed spectral variability. However, our study offers a more quantitative assessment of these time-variable atmospheric parameters.

In a complementary analysis, \citet{nasedkin2025jwst} performed independent retrievals on 24 phase-binned spectra from the same 0.8–11 $\rm \mu m$ dataset. Their results suggest that the variability is primarily driven by changes in effective temperature, as well as abundance variations in $\rm CO_{2}$ and $\rm H_{2}S$. They report a $\sim$ 250 K thermal inversion above 10 mbar across all rotational phases and propose that this inversion may be caused by auroral energy deposition into the upper atmosphere. No significant variability is found in any cloud property, including the patchy cloud fraction, vertical distribution, or optical depth.
Our interpretation agrees with their findings: we also retrieve largely invariant $\rm Mg_{2}SiO_{4}$ slab cloud properties, including the cloud fraction and scattering characteristics. However, our model identifies rotationally varying vertical distribution in the iron deck cloud and $\rm CO$ abundance. This discrepancy may stem from differences in wavelength coverage between the two studies. Our NIRISS/SOSS near-infrared data are not sensitive to the upper atmosphere (above 10 mbar) or to the dominant absorption features of $\rm CO_{2}$ and $\rm H_{2}S$. Nonetheless, the \texttt{Tempawral} demonstrates its capacity to retrieve complex thermal, chemical, and cloud structure variability, even from limited spectral bandpasses. The \hbox{$\sim$300\,K} perturbation near 1\,bar is likely indicative of localized hot spots, potentially driven by non-equilibrium $\rm CO/CH_{4}$ chemistry.

Broadband spectral time-series observations are essential to developing a more complete, three-dimensional understanding of brown dwarf atmospheres across a wider range of pressure levels. These would help disentangle the physical processes governing variability as a function of altitude. Notably, SIMP\,0136 is a known auroral emitter, exhibiting pulsed radio bursts \citep{kao2016auroral,kao2018strongest}. Auroral processes can contribute to photometric variability in both optical and infrared wavelengths, as downward-propagating electron beams modify local atmospheric temperature and opacity \citep{harding2013periodic}. For instance, energetic electrons colliding with atmospheric particles can cause localized heating, and the increased electron number density contributes excess free electrons, leading to an increase in the $\rm H^{-}$ population. Therefore, pairing \texttt{Tempawral} with future broadband time-series spectroscopy particularly spanning optical to mid-infrared wavelengths could offer an excellent opportunity to investigate auroral heating effects and aurora-induced chemistry in the upper atmosphere of SIMP\,0136. Such an approach would also allow for the full disentanglement of vertical structure, phase-dependent chemistry, and external energy deposition in the auroral brown dwarf.

\subsection{Note on Time-Series Model Parameterizations}


We note that \texttt{Tempawral} is highly sensitive to the choice of model parameterizations. Specifically, whether a given parameter is included as time-variable in the retrieval, and how its evolution across the rotational phase is modeled, can significantly influence the outcome. In our best-fit model, we included a temperature perturbation along with chemistry and cloud properties as time-varying parameters. This choice was motivated by the likelihood of disequilibrium chemistry creating hot spots. The variability pattern likely arises from a complex combination of mechanisms including patchy clouds, temperature perturbations, and chemical changes.

Figure~\ref{fig:data_eigenfit} presents a comparison case in which the retrieval was run without temperature perturbation. While changes in chemistry and clouds alone can reproduce the overall trend in the eigen-spectra, they significantly deviate in molecular absorption bands. This is because the missing temperature component fails to capture variations in the photosphere's location at different wavelengths. Three-dimensional general circulation models (e.g., \citealp{lee2023dynamically,lee2024dynamically}) suggest a highly complex interplay between clouds, chemistry, and the atmospheric temperature structure, leading to latitudinal variation and time-dependent storms at photospheric pressures. Brown dwarf variability likely arises from these intricate feedbacks. Therefore, including as many relevant time-varying parameters as feasible is essential to understanding how chemistry, clouds, and energy transport evolve coherently. Although such inclusion increases the risk of parameter degeneracies, these degeneracies may be less severe in eigen-spectra space. As shown in Figure~\ref{fig:sens}, variations in individual model parameters tend to produce distinct responses in the eigen-spectra.

The temperature perturbation used in this study assumes a symmetric shape centered on a reference pressure, intended to mimic hot or cold surface features. However, in the case of auroral heating, which is believed to cause a temperature inversion in the upper atmosphere, the perturbation would likely be asymmetric, with more energy deposited at lower pressures and decaying with depth. We did not adopt this asymmetric perturbation in the current work because our near-infrared data lacks sensitivity to altitudes higher than 10$^{-2}$ bar. As indicated by the dashed black curve in Figure~\ref{fig:data_pt_evole}, the temperature profile retrieved from the time-averaged spectrum appears isothermal in the upper atmosphere, unlike the temperature inversion retrieved from a broader wavelength dataset (e.g., NIRSpec + MIRI LRS; \citealp{nasedkin2025jwst}). This isothermal feature may reflect, to some extent, modulation due to auroral heating compared to the steeper temperature gradient expected in the absence of an additional upper-atmosphere heating source. Testing \texttt{Tempawral} with an asymmetric temperature perturbation using both near-infrared and broader coverage datasets like NIRSpec + MIRI LRS will be crucial to examine how inferred variability origins shift with wavelength coverage.

Another important finding is the inability to perfectly fit both the first and second eigen-spectra components simultaneously. In particular, the model fails to reproduce the second component in the water absorption band. This mismatch suggests the presence of multiple asynchronous water features across the atmosphere that a single sinusoidal variation with one phase-shift and variance cannot capture. To better approximate such complex distributions, two alternative time-variable time-distribution parameterizations could be explored in future work: 1) Multi-component sinusoidal modeling: Assigning multiple sinusoidal variation functions to a single parameter to represent spatially distinct patterns; 2) Fourier series expansion: Using a finite set of harmonic components to represent an arbitrary periodic distribution for each time-variable. These parameterizations could offer a more realistic representation of the heterogeneous and dynamic atmospheres of brown dwarfs.

\subsection{Caveat}

\texttt{Tempawral} offers a generalized time-resolved framework to probe the origins of spectroscopic variability and to gain insights into the atmospheric dynamics of brown dwarfs and exoplanets. However, two key assumptions underlying the model should be highlighted.

\subsubsection{Dependence on Baseline Retrieval}

In its forward modeling framework, \texttt{Tempawral} starts from a set of reference atmospheric parameters derived from the retrieval of the time-averaged spectrum. These reference values serve as the “first guess” of the atmosphere, upon which perturbations are applied to simulate time-variable spectra. As a result, the choice of baseline model from the time-averaged retrieval critically impacts the subsequent eigen-spectra retrieval. For ultra-cool brown dwarfs, the interplay between temperature structure, disequilibrium chemistry, and cloud properties can produce significant degeneracies in spectral retrievals \citep{burrows1997nongray, allard2001limiting, cooper2003modeling, burrows2011dependence, helling2014atmospheres, noll1997detection, oppenheimer1998spectrum, saumon2000molecular, geballe2009spectroscopic,wang2022unveiling}. In many cases, a given spectrum may be explained by multiple combinations of thermal profiles, cloud compositions and distributions, and chemical abundances. While model selection based on the Bayesian Information Criterion (BIC) or Bayesian evidence is standard practice, comparisons can be biased when models use different parameterizations or prior ranges. Consequently, any artifact introduced in the baseline model may propagate through and bias the interpretation of variability and atmospheric dynamics. To ensure robust conclusions, it is essential to compare eigen-spectra retrievals based on different baseline models. Conversely, the rich information contained in time-series data can aid in ruling out baseline models that poorly capture the variability, providing a feedback loop to improve time-averaged retrievals.

\subsubsection{Sinusoidal Assumption of Time-Variable Evolution}\label{sec:Sinusoidal}

We also note a limitation in assuming that time-varying parameters follow sinusoidal evolution over the rotation. This is a reasonable approximation when variability arises from discrete, rotating surface features such as chemical inhomogeneities or hot spots. As such features rotate in and out of view, the corresponding time-variable, e.g., gas abundance or longitudinally-averaged temperature can naturally follow a sinusoidal pattern. However, this assumption may break down for more spatially extended features such as global-scale patchy clouds or rapidly evolving phenomena like storms or auroral pulses. These processes may produce irregular or asymmetric time-series patterns that cannot be captured by a single sinusoid. More sophisticated parameterizations such as a combination of multiple sinusoids or stochastic models would be needed to realistically model such dynamics. That said, for SIMP\,0136, these rapidly evolving processes do not appear to be the dominant source of variability. Independent retrievals by \citet{nasedkin2025jwst} indicate that several dominant time-variable parameters, such as effective temperature and the abundances of $\rm CO_{2}$ and $\rm H_{2}S$ can exhibit nearly sinusoidal variations over the rotation period. In Figure~\ref{fig:ait_score}, we show that the temporal contributions of the first two PCA components exhibit sinusoidal-like behavior, consistent with slowly evolving or rotationally modulated features. This supports the use of sinusoidal distributions for the dominant time-variables in this particular case. Finally, we note that the top-of-atmosphere flux from an averaged 1D profile is not equivalent to the disk-averaged flux of a heterogeneous surface. Modeling disk-averaged spectra with a single thermal and chemical profile already introduces some bias, and the impact of further assuming sinusoidal variations in time-varying parameters should ultimately be assessed with more sophisticated 3D GCMs.

\section{Summary}
\label{sec:sum}

In this paper, we introduce a novel time-resolved retrieval framework, \texttt{Tempawral}, designed to directly infer the physical drivers of atmospheric variability from time-series spectra. \texttt{Tempawral} operates on eigen-spectra derived from the full time-series dataset, offering several advantages: it avoids the attenuation of variability patterns caused by spectral binning, utilizes most of spectral variability information simultaneously, and enables direct, quantitative estimates of the contribution from each physical parameter. The framework begins with a baseline retrieval on the time-averaged spectrum of the observed object to obtain a first-principles estimate of its atmospheric state. Time-variable perturbations are then applied to this reference state to reproduce the variability patterns observed in the time-series data. This approach provides a generalized framework for modeling the temporal evolution of brown dwarf and exoplanet atmospheres across different effective temperatures and surface gravities.

We validate the framework using mock time-series simulations of a SIMP\,0136-like brown dwarf and then apply it to real observations from the JWST/NIRISS-SOSS instrument. The key findings of this work are summarized below:

1. Mock simulations showcase \texttt{Tempawral}'s ability to recover variability driven by heterogeneous cloud coverage, chemical abundances, and temperature structure. The method successfully isolates each effect through eigen-spectra decomposition.

2. Reliable constraints on the variance of time-varying parameters require a signal-to-noise ratio $\ge$ 100 and $\gtrsim$ 10 evenly spaced samples across the rotational phase. This sampling strategy strikes a balance between capturing the variability in the data and maintaining computational feasibility.
The total computational time for eigen-spectra retrieval on a time-series data set is comparable to that of a traditional single-snapshot spectral retrieval.

3. Phase shifts between different time-dependent parameters (e.g., chemistry, cloud properties, temperature structure) introduce strong degeneracies in eigen-spectra retrievals. Including flux variability amplitude information in the likelihood function improves constraints, but fully disentangling phase shifts remains challenging. Observations covering multiple rotational periods are likely required to resolve these issues.

4. When applied to JWST/NIRISS near-infrared time-series data of SIMP\,0136, the variability is best explained by a $\sim$300\,K temperature perturbation near 1 bar, combined with changing abundances of $\rm H_{2}O$, $\rm CO$, $\rm FeH$, and variations in the thickness of the iron cloud deck. No significant variability is found in the properties of the slab $\rm Mg_{2}SiO_{4}$ cloud, including cloud fraction and scattering properties, while the iron deck cloud distribution varies as a function of phase. A possible temperature inversion driven by auroral heating or aurora-induced chemical changes is not detected in this analysis, likely due to the limited sensitivity of near-infrared data to these upper-atmospheric processes.

As of Cycle 3, JWST has performed $\sim$150\,hours of time-series spectroscopy monitoring on L, T, and Y dwarfs. This work establishes a generalized and flexible retrieval framework to interpret such data, enabling deeper insight into the complex atmospheric dynamics, chemical variability, and energy transport in substellar atmospheres. For future work, the viewing geometry is a critical factor in understanding variability patterns in brown dwarf atmospheres. Incorporating this geometric context into atmospheric models, together with more sophisticated time-dependent parameterizations capable of capturing multiple asynchronous surface features across different regions, will be an important next step. On the other hand, \texttt{Tempawral} performs retrievals directly on the eigen-spectra, which correspond to eigenvectors capturing the dominant variance in the time-series spectra. The PCA decomposition also encodes the relative contributions of each component to the observed variability. This information should be reconsidered and incorporated into the current framework to enable more accurate reconstructions of atmospheric variability maps.

\section*{Acknowledgements}
FW and BB acknowledge support from UK Research and Innovation Science and Technology Facilities Council [ST/X001091/1].  JMV acknowledges from a Royal Society - Research Ireland University Research Fellowship (URF/1/221932).

\section*{Data Availability}
All data underlying this article are publicly available upon reasonable request to the corresponding author.



\bibliographystyle{mnras}
\bibliography{ref} 




\appendix

\section{PCA EXPLAINED VARIANCE}

\begin{figure}
	\includegraphics[width=1\columnwidth]{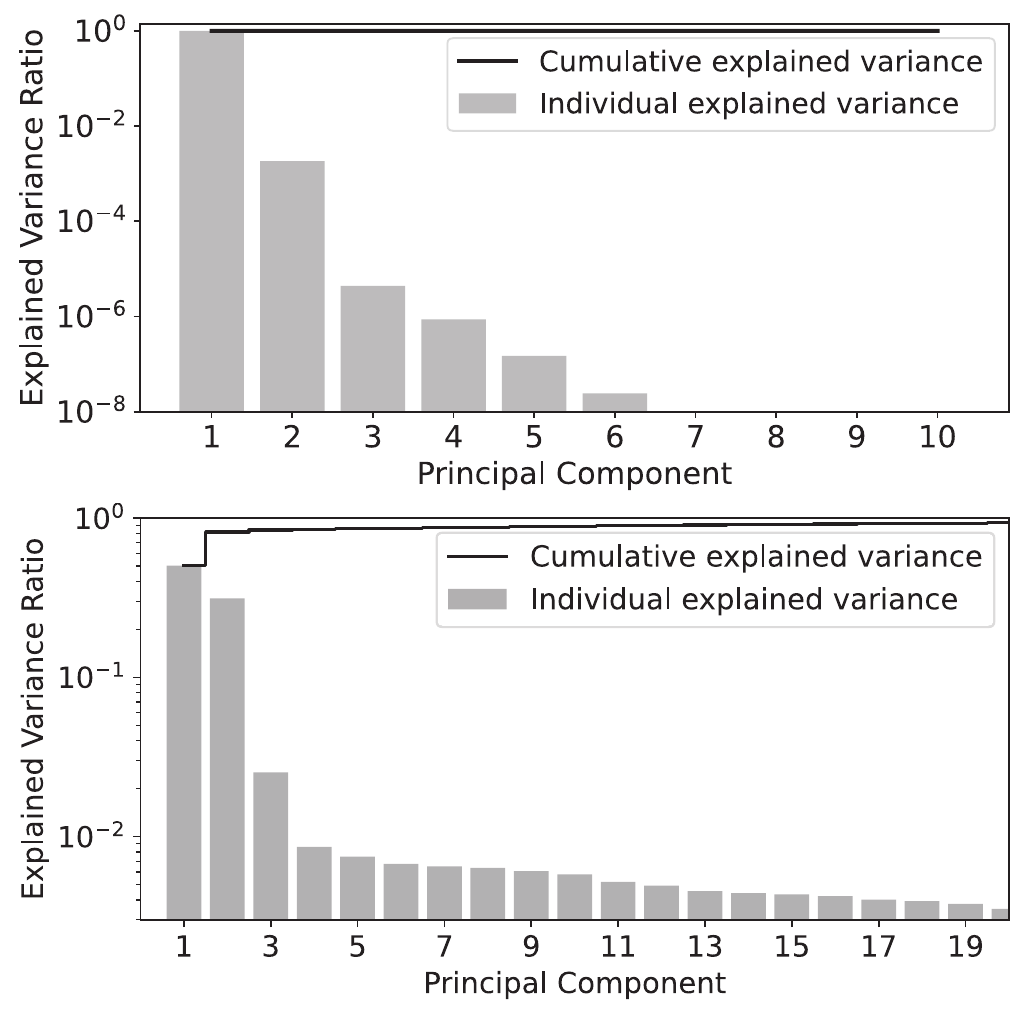}
    \caption{Individual explained variance for each of the principal components shown as a histogram, and cumulative explained variance shown with solid lines. \textbf{Upper panel:} The first two components explain $\ge 99\%$ the variance in canonical simulated model data (Section~\ref{subsec:Eigen}). \textbf{Bottom panel:} The first two components explain $\ge 81\%$ of the variance of JWST SIMP~0136 data (Section~\ref{sec:data}).}
    \label{fig:explained_ratio}
\end{figure}

Figure~\ref{fig:explained_ratio} shows the explained variance for each component in the cases of our canonical simulated model data (Section~\ref{subsec:Eigen}) and JWST SIMP~0136 data (Section~\ref{sec:data}), the first two components explain $\ge 99\%$ and $\ge 81\%$ of the variance, respectively.

\section{Corner Plot}

Retrieved posterior distributions of model parameters obtained in both the simulated time-series data (Section~\ref{sec:result}) and the JWST SIMP~0136 observations (Section~\ref{sec:application}) using \texttt{Tempawral}.

\begin{figure*}
	\includegraphics[width=2.1\columnwidth]{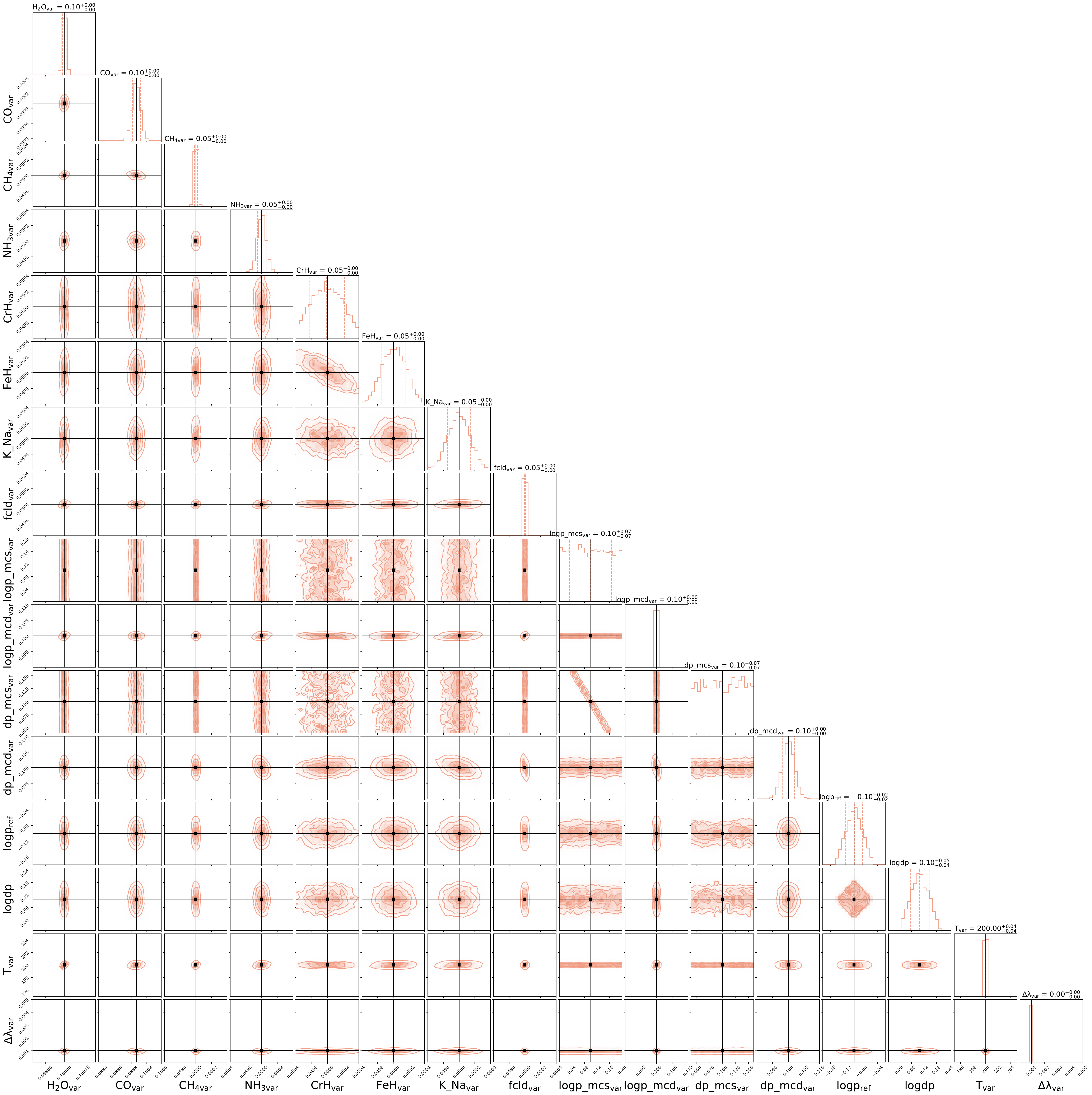}
    \caption{Corner plot summarizing the posterior probability distributions for our canonical model. The blue solid line in each histogram represents the true value of the parameter in the forward model (Table~\ref{tab2}). The black dashed lines mark the 16\%, 50\%, and 84\% quantiles. On top of each histogram is the retrieved median value and $\pm 1\sigma$ range.}
    \label{fig:corner_sim}
\end{figure*}

\begin{figure*}
	\includegraphics[width=2.1\columnwidth]{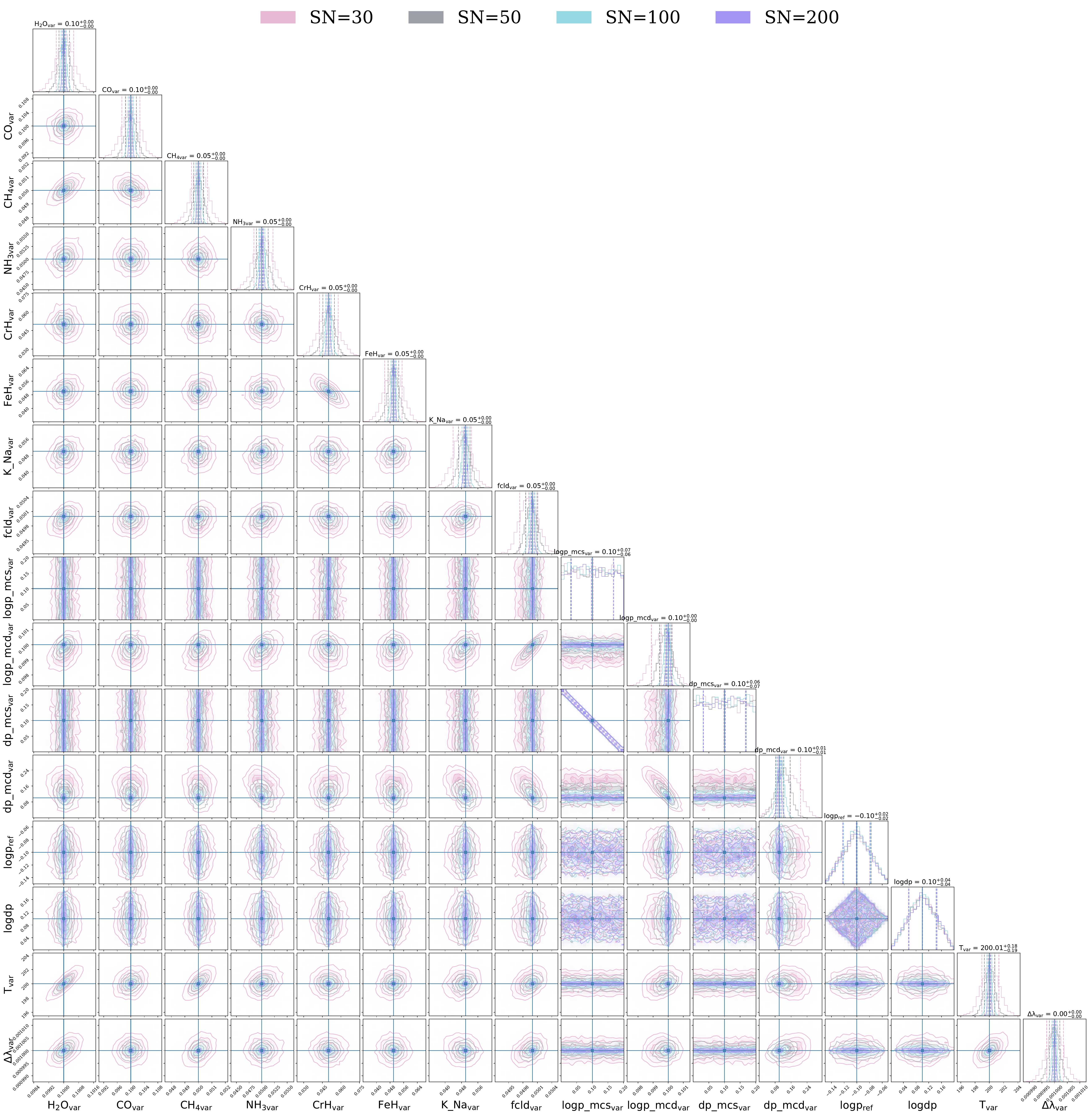}
    \caption{Corner plot summarizing the posterior probability distributions under different noise levels of 30, 50, 100, 200. The blue solid line in each histogram represents the true value of the parameter in the forward model (Table~\ref{tab2}). On top of each histogram is the retrieved median value and $\pm 1\sigma$ range.}
    \label{fig:corner_SNs}
\end{figure*}

\begin{figure*}
	\includegraphics[width=2.1\columnwidth]{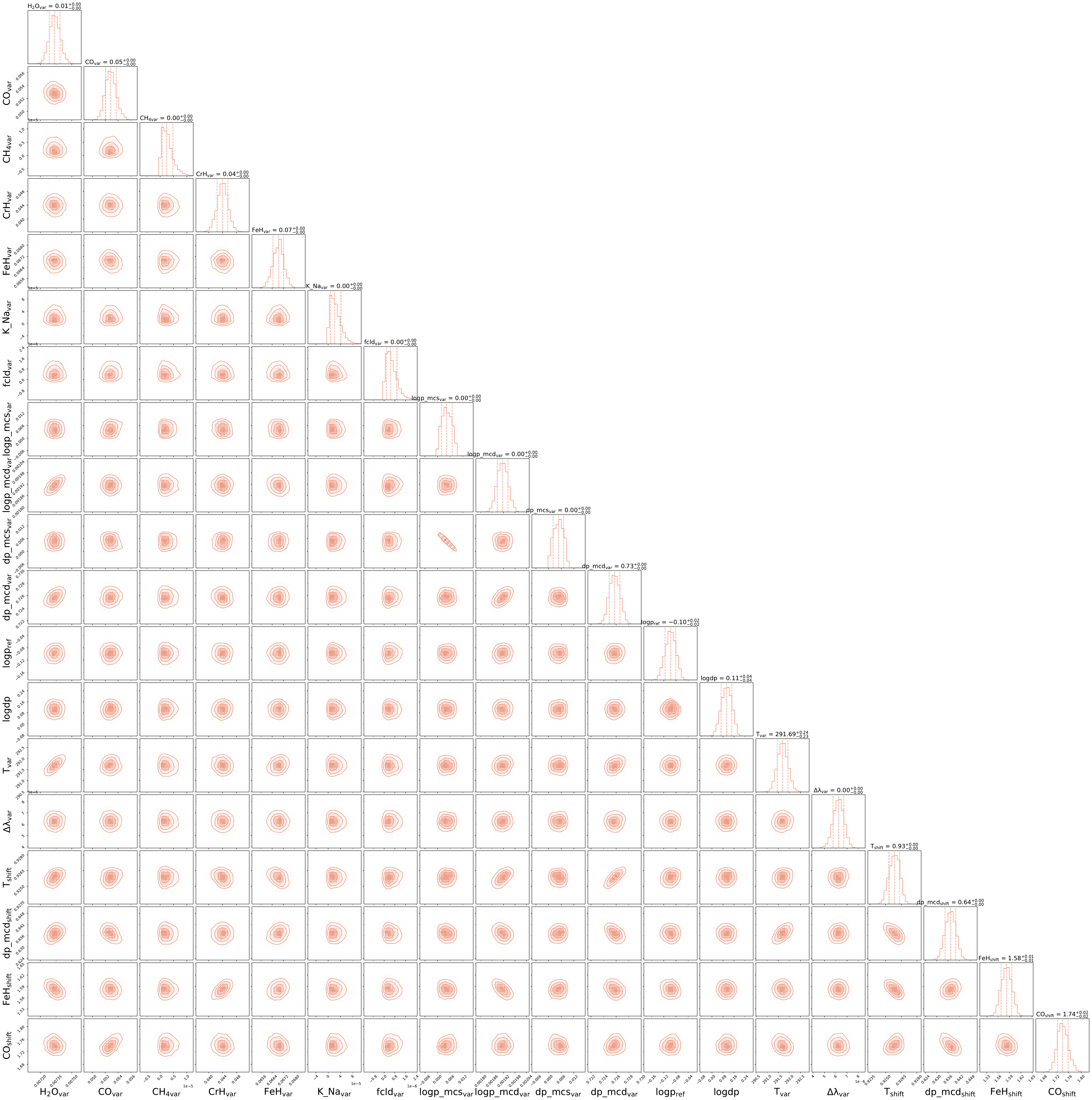}
    \caption{Corner plot summarizing the posterior probability distributions of each time-variables in SIMP\,0136 data. On top of each histogram is the retrieved median value and $\pm 1\sigma$ range.}
    \label{fig:corner_data}
\end{figure*}

\section{SIMP\,0136 Reconstructed Light Curve}

Reconstructed light curve across five photometric bands derived from the post-processed results of Figure~\ref{fig:data_tspec_fit}, to further demonstrate the fractional variability recovered by \texttt{Tempawral}.

\begin{figure*}
	\includegraphics[width=2.1\columnwidth]{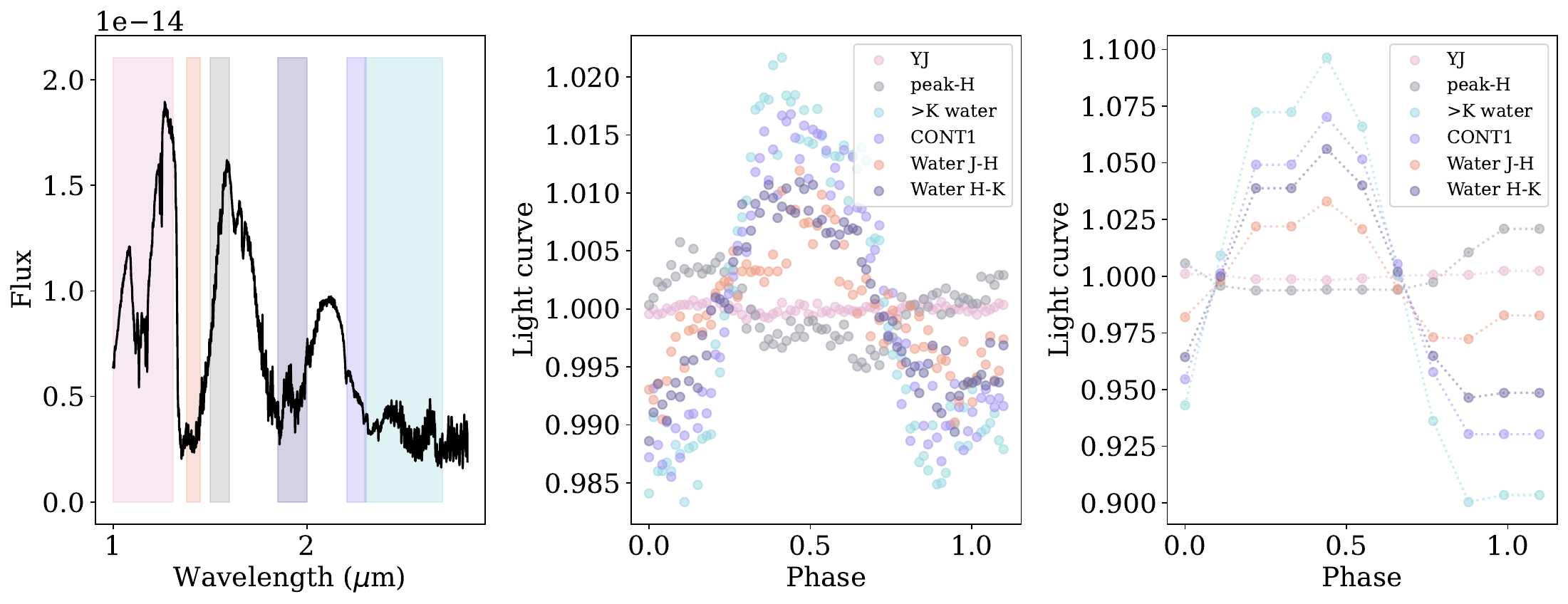}
    \caption{
    \textbf{Left:} Time-averaged spectrum of SIMP\,0136 over a full rotation with NIRISS/SOSS. Colored bands indicate the '$\rm YJ$', '$\rm peak$-$\rm H$', '$\rm > K$ water', 'CONT1', 'Water $\rm J$–$\rm H$', and 'Water $\rm H$–$\rm K$' bands.
    \textbf{Middle:} Corresponding light curves for each spectral bin of the data.
    \textbf{Right:} Light curves reconstructed from the best-fit variability maps shown in Figure~\ref{fig:data_tspec_fit}.
    The recovered light curves are slightly offset relative to the data as they are normalized by the data uncertainties. While the phase resolution is limited by the use of only 10 sampling bins in \texttt{Tempawral}, the recovered variability patterns in each spectral band remain consistent with the observed light curves.}
    \label{fig:lightcurve}
\end{figure*}

\section{Phase-dependent contributions of SIMP\,0136 Eigen-spectra}

\begin{figure*}
	\includegraphics[width=2.1\columnwidth]{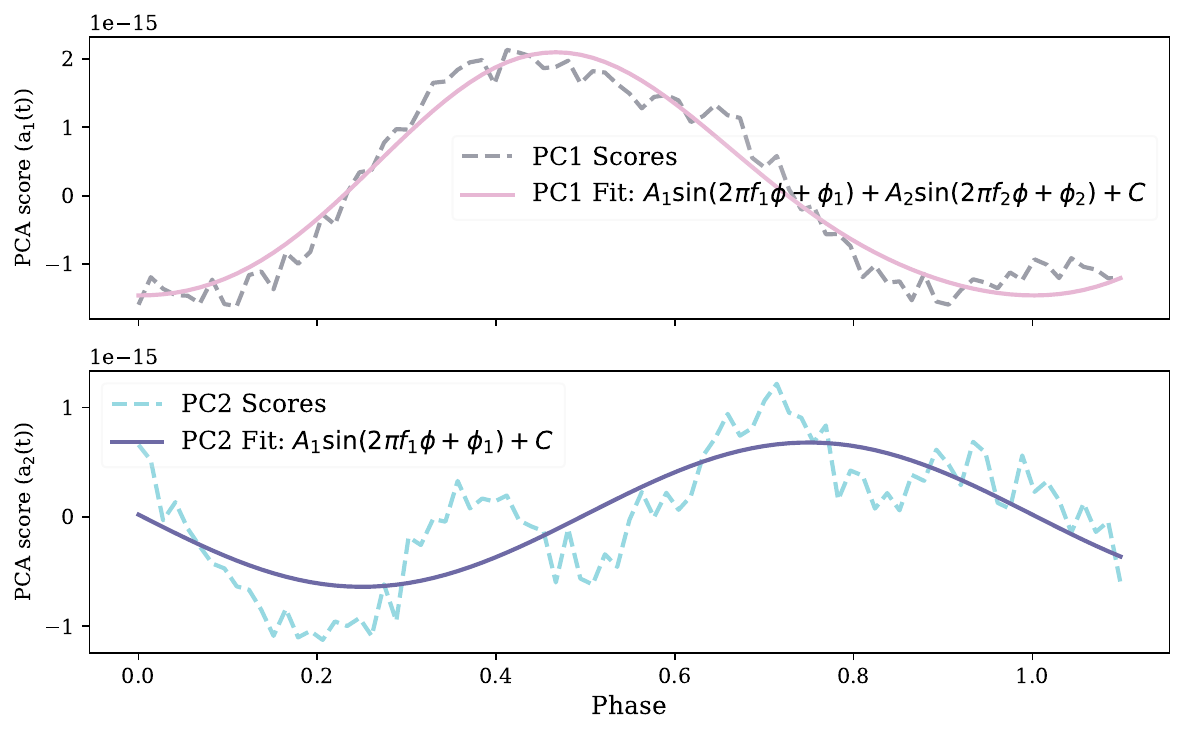}
    \caption{Phase-dependent contribution scores ($a_{i}(t)$) of the first two principal component of SIMP\,0136 data. These contribution scores quantify how strongly the $i$-th eigen-spectrum contributes to the time-series spectrum at time $t$.
    \textbf{Top}: PC1 is fitted with a two-term harmonic sinusoidal model. The fitted curve represents the sum of two sine functions with fundamental and second harmonic frequencies, capturing the two main variability drivers in the spectral data. \textbf{Bottom}: PC2 is fitted with a single sinusoidal model.}
    \label{fig:ait_score}
\end{figure*}

Figure~\ref{fig:ait_score} show that the phase-dependent contribution scores ($a_{i}(t)$) of the first two principal component of SIMP\,0136 data, which exhibit sinusoidal-like behavior, consistent with slowly evolving or rotationally modulated features. This supports the use of sinusoidal distributions for the dominant time-variables in this work.

\bsp	
\label{lastpage}
\end{document}